\documentclass[aps,prb,showpacs,twocolumn,superscriptaddress]{revtex4-1}
\usepackage{amsmath}
\usepackage{amssymb,amscd,bm,dsfont,wasysym,mathrsfs,latexsym,psfrag,accents,mathtools}
\usepackage{graphicx}
\usepackage{multirow}
\usepackage{dcolumn}
\usepackage{float}
\usepackage{color}
\usepackage{xcolor}
\usepackage{comment}

			\newcommand{\e}[1]{\begin{align}{#1}\end{align}}	
			\newcommand{\es}[1]{\begin{align*}{#1}\end{align*}}	
			\newcommand{\m}[1]{\begin{multline}{#1}\end{multline}}	
		
		\newcommand{\f}[2]{\frac{#1}{#2}}

		\newcommand{\la}[1]{\label{#1}}

		\newcommand{\q}[1]{Eq.\ (\ref{#1})}
		\newcommand{\qq}[2]{Eqs.\ (\ref{#1}-\ref{#2})}
		\newcommand{\s}[1]{Sec.\ \ref{#1}}
		\newcommand{\fig}[1]{Fig.\ \ref{#1}.}		
			
		\newcommand{\app}[1]{App.\ \ref{#1}}
		\newcommand{\ocite}[1]{Ref.\ \onlinecite{#1}}
		
		\newcommand{\iwith}{\ins{with}}

		\newcommand{\eq}{=&\;}

		\newcommand{\R}{\mathbb{R}}
		
		\newcommand{\C}{\mathbb{C}}

	\newcommand{\eikr}{e^{i\bk \cdot \br}}

\newcommand{\nabk}{\nabla_{\boldsymbol{k}}}

\usepackage{graphicx}
\usepackage{float} 
\usepackage{braket} 
\usepackage{bbold}
\usepackage[ansinew]{inputenc} 
\usepackage{hyperref}
\usepackage{color}
\usepackage{comment}
\usepackage{amsthm}

\newcommand{\var}{\varepsilon}

\newcommand\as{\;\;\;\;}

\newcommand{\bk}{\boldsymbol{k}}

\newcommand{\br}{\boldsymbol{r}}

\newcommand{\bt}{\boldsymbol{t}}

\newcommand{\bA}{\boldsymbol{A}}

\newcommand{\bG}{\boldsymbol{G}}

\newcommand{\bR}{\boldsymbol{R}}
\newcommand{\bS}{\boldsymbol{S}}

\newcommand{\bze}{\boldsymbol{0}}

\newcommand{\bDelta}{\boldsymbol{\Delta}}

\newcommand{\bvarpi}{\boldsymbol{\varpi}}

\newcommand{\sx}{\sigma_{\sma{1}}}

\newcommand{\sz}{\sigma_{\sma{3}}}

\newcommand{\ins}[1]{\;\;\;\text{#1}\;\;\;}

\newcommand{\cali}{{\cal I}}

\newcommand{\calp}{{\cal P}}
\newcommand{\calq}{{\cal Q}}

\newcommand{\mo}{\text{-}1}

\newcommand{\minus}{\text{-}}

\newcommand{\braopket}[3]{\big\langle #1 \big| #2 \big| #3 \big\rangle}

\newcommand{\pdg}[1]{{#1}^{\phantom{\dagger}}}

\newcommand{\lin}{\notag \\}

\newcommand{\ab}{\alpha\beta}

\newcommand{\bpm}{\begin{pmatrix}}
\newcommand{\epm}{\end{pmatrix}}

\newcommand{\sma}[1]{\scriptscriptstyle{#1}}

\newcommand{\Z}{\mathbb{Z}}

\newcommand{\nocontentsline}[3]{}
\newcommand{\tocless}[2]{\bgroup\let\addcontentsline=\nocontentsline#1{#2}\egroup}

\newtheorem{theorem}{Theorem}
\newtheorem*{theorem*}{Theorem} 
\newtheorem{definition}{Definition}
\newtheorem*{definition*}{Definition} 

\graphicspath{{./Figures/}}

\begin{document}

\title{No-go theorem for topological insulators and sure-fire recipe for Chern insulators}
\author{A. Alexandradinata} \author{J. H\"oller} \affiliation{Department of Physics, Yale University, New Haven, Connecticut 06520, USA} 
  

\begin{abstract}

For any symmorphic magnetic space group $G$, it is proven that topological band insulators with vanishing first Chern numbers cannot have a groundstate composed of a \emph{single}, energetically-isolated band. This no-go statement implies that the minimal dimension of  tight-binding Hamiltonians for such topological insulators is \emph{four} if the groundstate is stable to addition of trivial bands, and \emph{three} if the groundstate is unstable. A sure-fire recipe is provided to design models for Chern and unstable topological insulators by splitting elementary band representations; this recipe, combined with recently-constructed Bilbao tables on such representations, can be systematized for mass identification of topological materials. All results follow from our theorem which applies to any single, isolated energy band of a $G$-symmetric Schr\"odinger-type or tight-binding Hamiltonian: for such bands, being topologically trivial is equivalent to being a band representation of $G$.  
\end{abstract}
\date{\today}

\maketitle


A real-space-centric perspective on  topological band insulators is emerging from various directions,\cite{Resta1994,Soluyanov2011,Maryam2014,Read2017,Bradlyn2017,Cano2017,Holler2017} with the unifying theme that topological nontriviality  is fundamentally linked to an obstruction to constructing Wannier functions.\cite{Thouless1984,Thonhauser2006,Brouder2007,Budich2014,Read2017} In band insulators, the existence of Wannier functions  has traditionally justified that electrons form exponentially localized wavepackets in real space, and therefore the effects of local disturbances are short-ranged.\cite{Kohn1973,Cloizeaux1964,Nenciu1983} Despite the similarity of such localized wavepackets with the electronic orbitals of atoms, there remains a sharp, group-theoretic distinction between solids and a lattice of spatially-isolated {atoms} -- Wannier functions and atomic orbitals  may transform differently under crystallographic point-group symmetries that preserve at least one spatial point, as exemplified by rotations or time reversal. This distinction was first pointed out by Soluyanov and Vanderbilt\cite{Soluyanov2011} for the Kane-Mele\cite{kane2005A,Kane2005b,QSHE_bernevig,HgTe_bernevig,QSHE_Rahul,fu2006} topological insulator in Wigner-Dyson symmetry class AII:\cite{Dyson1962} in this phase it is not possible to construct a Kramers pair of  Wannier functions centered at the same spatial point,\cite{Soluyanov2011,Read2017,Bradlyn2017} as illustrated in \fig{fig:pie}(d). Alternatively stated, these Wannier functions do not locally represent time-reversal symmetry.

\begin{figure}
\centering
\includegraphics[width=8cm]{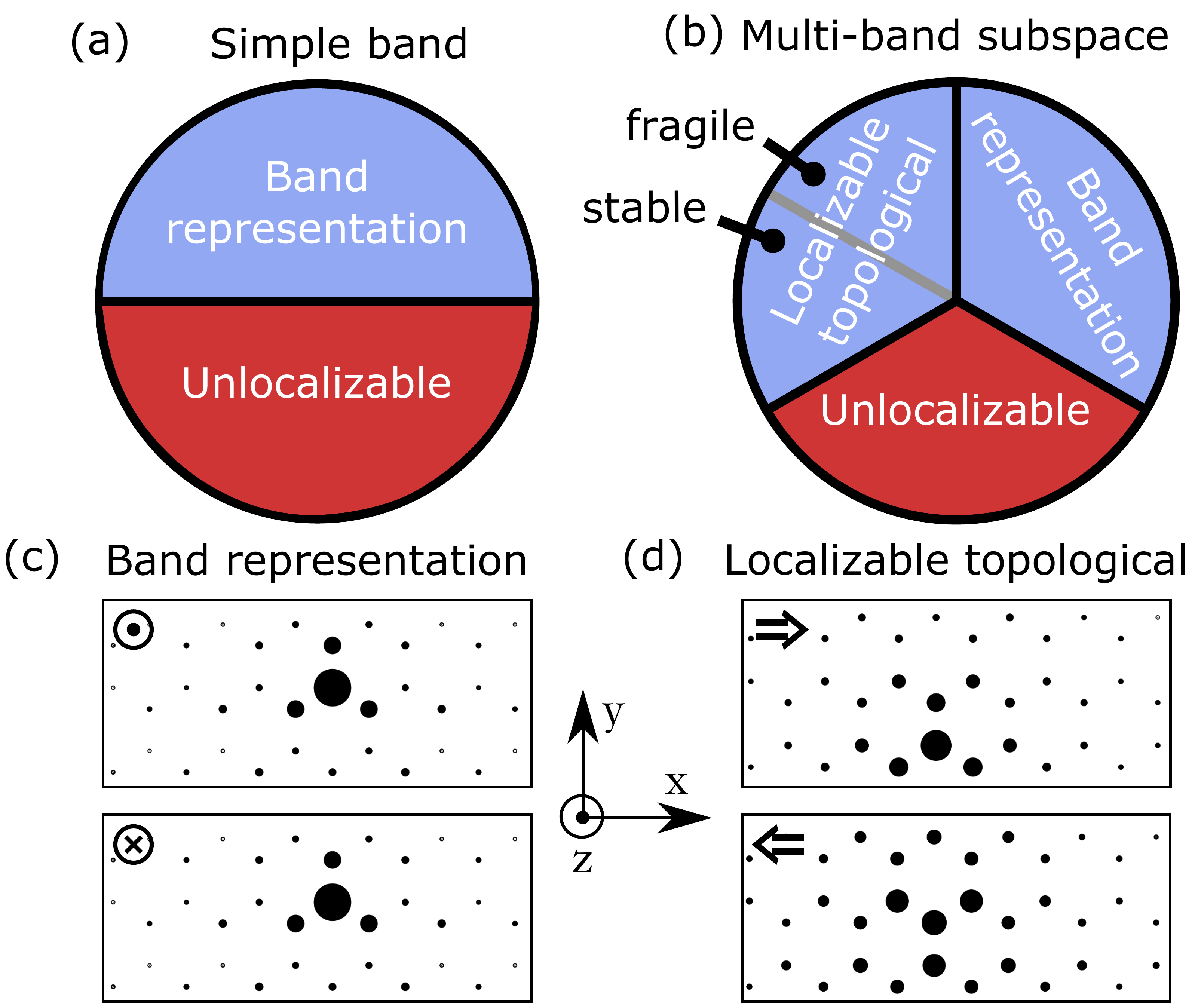}
\caption{ (a{-}b) Topological categorization of a simple band [(a)] and, complementarily, $(N{>}1)$-band subspaces [(b)]. In spatial dimension $d{\leq}3$, being localizable is equivalent to having vanishing first Chern class ($c_1$), as colored blue; phases with $c_1{\neq}0$ are colored red. (c{-}d) Comparison of Wannier functions in a trivial  and  Kane-Mele phase, with the symmetry of a honeycomb lattice with staggered sublattices.  The size of each dot measures the weight of the Wannier function on a lattice site; the expectation value of spin for each Wannier function is indicated by an arrow in the inset. (c) In the trivial phase, Wannier functions form Kramers pairs centered on the same lattice site. (d) In the Kane-Mele phase, every Wannier function is centered on a different lattice site.
\la{fig:pie} }
\end{figure}



In the Kane-Mele model, the bands which are filled at zero temperature exemplify a localizable topological band. By `localizable', we mean that the band is unitarily equivalent to a set of Wannier functions; by `localizable topological', we further impose that the Wannier functions  are obstructed from satisfying the following \emph{local symmetry condition}: for any spatial point $\bvarpi$, all Wannier functions centered at $\bvarpi$ form a representation of {all} point-group symmetries that preserve $\bvarpi$. If a localizable topological band is the filled band of an insulator, we refer to this insulator as a localizable topological insulator.\footnote{Our results do not immediately apply to band topological superconductors, which have an additional
particle-hole and/or chiral symmetries in the Bogoliubov-de-Gennes formalism.\cite{schnyder_classify3DTIandTSC} The possibility to formulate band representations for superconductors remains an open and interesting question.} Included in this category  are all $(d{\leq}3)$-dimensional topological  insulators  with vanishing first Chern class ($c_1{=}0$), and whose protective symmetries are classified by the magnetic space groups\cite{magnetic_groups} (numbering $1651$ in $d{=}3$). Both time-reversal-invariant\cite{Bradlyn2017} and magnetically-ordered\cite{Watanabe2017a} insulators are considered; in the former case the first Chern class vanishes by symmetry.\cite{Panati2007} Examples of localizable topological insulators that have materialized in laboratories include Bi$_2$Se$_3$\cite{zhang2009,hsieh2008} (a 3D $\Z_2$ topological insulator),\cite{fukanemele_3DTI,Inversion_Fu,moore2007,Rahul_3DTI} SnTe\cite{Hsieh_SnTe,Xu_observeSnTe,Tanaka_observeSnTe} and KHgSb\cite{Hourglass2016,Alexandradinata2016,Ma_discoverhourglass} (topological crystalline insulators).\cite{Fu2011,Liu2014,Alexandradinata2014g,Alexandradinata2014c,Shiozaki2014,Fang2015,Shiozaki2015,Shiozaki2016,Shiozaki2017,Classification_Chiu,AZ_mirror}

This work presents rigorous results that apply in any spatial dimension $d$ and to solids whose magnetic space groups (denoted $G$) are symmorphic, i.e., $G$ consists only of symmetries that are factorizable into products of point-group symmetries with lattice translations. A case in point is SnTe, whose rocksalt structure has the symmetry of the symmorphic space group 225; in contrast,   nonsymmorphic KHgSb is symmetric under glide, which is the product of a reflection with half a lattice translation.  

Our first result is a no-go statement: a localizable topological band cannot be  a simple band. By `simple', we mean a single band that is nondegenerate (in energy) throughout the Brillouin torus.  A localizable band that is unitarily equivalent to \emph{locally symmetric} Wannier functions is defined to be a band representation;\cite{Zak1979,Zak1981,Evarestov1984,Bacry1988}  colloquially, a band representation resembles a lattice of locally-symmetric atomic orbitals,\cite{Holler2017} as illustrated in \fig{fig:pie}(c).  The contrapositive of the no-go statement is that  any simple band that is not a band representation cannot be localizable, i.e., it has nontrivial first Chern class ($c_1{\neq}0$); this is presented pictorially in \fig{fig:pie}(a-b), which compares the topological categorization of simple bands with non-simple bands.\footnote{A non-simple band necessarily has rank $N{>}1$, but we do not require that a $(N{>}1)$-band subspace is connected by band touchings.
Our definition of  unlocalizable and localizable topological bands [cf.\ \fig{fig:pie}(a-b)] should in principle coincide with that of `topologically non-trivial bands' in \ocite{Bradlyn2017}, which `cannot be continued to any atomic limit without either closing a gap or breaking a symmetry.' Note that they have not defined topological triviality in the category of complex vector bundles, as we have done in this work. To verify their notion of non-triviality, it is necessary to construct a continuous family of Hamiltonians that symmetrically interpolates between the Hamiltonian of interest to an atomic-limit Hamiltonian (which is not uniquely defined). In comparison, our definition offers a diagnostic test (for the non-existence of locally-symmetric Wannier functions) that can be implemented on a single Hamiltonian of interest.} Our results  follow from a theorem that applies to simple bands occurring as energy eigenfunctions of Schr\"odinger-type or tight-binding Hamiltonians with the symmetry $G$: for such simple bands, being topologically trivial (in the category of complex vector bundles) is equivalent to being a band representation of $G$.
 
The standard notion of topological triviality, as well as its relation to the existence of Wannier functions, is reviewed in  \s{sec:triviality} to further motivate the theorem. A more general and precise statement of the theorem is provided in \s{sec:theorem}.  One may question if there is a loss of generality in our restriction to symmorphic $G$. Actually the hypothesis of the theorem becomes superfluous for all nonsymmorphic magnetic space groups
with at least one nonsymmorphic element (a symmetry that is partially a translation by a fraction of a lattice vector, e.g., screw or glide), and possibly even for the minority of nonsymmorphic groups without nonsymmorphic elements; these groups just do not allow for simple bands.\footnote{For space groups with at least one nonsymmorphic element, 
bands must be nontrivially connected as a graph due to the monodromy of symmetry representations, as proven in the supplementary material of \ocite{Holler2017}. Less general proofs exist for solids without spin-orbit coupling in \ocite{Parameswaran2013} and for band representations only in \ocite{Michel1999}. All type I (without time-reversal symmetry) and II (with time-reversal symmetry by itself) nonsymmorphic magnetic space groups have nonsymmorphic elements in $d{=}2$, but not in $d{\geq}3$. For nonsymmorphic space groups without nonsymmorphic elements, we are not aware of a general proof that simple bands do not exist; however this seems empirically to be true for specific case studies in $d{=}3$.\cite{Michel1999,Parameswaran2013,Po2016,Watanabe2017a}}


  Applications of our no-go statement include: (i) the establishment of a minimal dimension (at each wavevector) for the tight-binding Hamiltonian of any localizable topological insulator with symmorphic symmetry. In the real-space perspective, the dimension of a tight-binding Hamiltonian is the number of orthogonal Wannier functions  that span the tight-binding Hilbert space in one unit cell.
   The minimal dimension  depends on whether the localizable topological groundstate is stable, i.e., whether it remains localizable topological upon addition of band representations -- stable phases  are classified by topological K-theory,\cite{Freed2013,Shiozaki2017}  and unstable (or `fragile'\cite{Po}\footnote{`Fragility' is a symmetry-enriched generalization of the notion of `stable triviality' in bundle theory.\cite{Hatcher2009,Read2017}   Nontriviality of the fragile topological phases may manifest as a spectral flow in the holonomy over noncontractible Brillouin-zone loops.\cite{Alexandradinata2014c,Cano2017a}}) phases manifest in a finer classification of vector bundles.\cite{DeNittis2014,DeNittis2017} We find that the minimal dimension is four in the stable case, and three in the fragile case. Additionally, we propose (ii) a  sure-fire recipe to design and/or identify  bands with nontrivial Chern number, as well as fragile localizable topological bands -- this recipe may be systematized  for mass identification of topological materials. (i-ii) and other applications are elaborated in \s{sec:applications}. 


\s{sec:generalizeSOC} generalizes the theorem to half-integer-spin representations of point-group symmetries, with application to  solids with spin-orbit coupling.  After establishing a few preliminary results on space groups and their representations in \s{sec:preliminaries}, we prove in \s{sec:analytictrivial} that a simple band that is a band representation is topologically trivial, and in \s{sec:forwardarrow} that a topologically trivial simple band is a band representation. Finally in \s{sec:discussion}, we summarize our results from the perspective of establishing rank constraints for bands.

\section{Existence of Wannier functions for topologically trivial bands}\la{sec:triviality}

Our theorem may be viewed as a symmetry-refined analog of known relations between topological nontriviality and an obstruction to the existence of Wannier functions; these relations will be briefly reviewed in \s{sec:trans}, and subsequently in \s{sec:pointgroup} we will introduce crystallographic point-group symmetry to further motivate our theorem.



\subsection{Insulators with discrete translational symmetry}\la{sec:trans}

Let us first consider $d$-dimensional band insulators in Wigner-Dyson symmetry class A, with the additional symmetry of discrete translations in $d$ independent directions. A well-recognized form of topological nontriviality arises for two-dimensional insulators whose  Hall conductance $C_1 e^2/h$ is quantized in units of fundamental constants; $C_1$ is the first Chern number (also known as the Thouless-Kohmoto-Nightingale-den Nijs\cite{Thouless1982}invariant) of the filled bands. Generally for $d$-dimensional solids, a nonzero $C_1$ in any two-dimensional closed submanifold of the Brillouin $d$-torus ($T^d$) is equivalent to a nonzero first Chern class ($c_1{\neq}0$).\cite{Panati2007} The foundational works of Nenciu,\cite{Nenciu1983} Panati\cite{Panati2007} and Brouder et al.\cite{Brouder2007} have culminated in an equivalence between the vanishing of the first Chern class ($c_1{=}0$) and the existence of Wannier functions (i.e., localizability) in $(d{\leq}3)$-dimensional solids. This equivalence broadly  applies to $N$-band subspaces for any $N{\geq}1$, including the case of simple bands ($N{=}1$); we shall refer to $N$ as the rank.

For $d{\leq}3$, the vanishing of the first Chern class ($c_1{=}0$) is equivalent to the band subspace being topologically trivial.\cite{DeNittis2017,Peterson1959} Throughout this letter, we adopt the standard definition of topological triviality from the theory of vector bundles.\cite{Husemoller1994} Applied to the Bloch problem, an $N$-band subspace is topologically trivial if there exist Bloch functions\footnote{Bloch functions do not have to correspond to energy bands; sometimes these are referred as quasi-Bloch functions.\cite{Nenciu1991,Brouder2007,Panati2007}} which  span the $N$-dimensional vector space at each quasimomentum ($\bk$), and are continuous and periodic over the Brillouin $d$-torus $T^d$.  To translate between band- and bundle-theoretic languages,  the set of $N$ Bloch functions form an $N$-dimensional vector space at each $\bk{\in}T^d$; the union of all such vector spaces over the base space $T^d$ defines a rank-$N$ complex vector bundle. If there exist  $N$ (continuous and periodic) sections that span the $N$-dimensional vector space at each $\bk$, we say that this bundle is topologically trivial in the category of complex vector bundles. We will interchangeably use `sections' with `Bloch functions', and `vector bundles' with `bands'. For most of this work, `topological triviality' should implicitly be understood as for complex vector bundles, though we shall remark briefly on real vector bundles in \s{sec:generalizeinversion}. 

The above definition of triviality applies to any $d$. Especially for $d{\geq}4$, which is physically realizable in cold-atomic\cite{Price2015,Lohse2018} and electrical circuits\cite{Ningyuan2015,Albert2015,Lee2018}, being topologically trivial more stringently constrains the band subspace than having a trivial first Chern class -- additional constraints include (but is not exhausted by)\cite{Hatcher2009,DeNittis2014} the triviality of all higher Chern classes. However, a nontrivial higher Chern class can only be realized by multi-band subspaces; in the absence of crystallographic symmetry, simple bands are fully classified by the first Chern class in any dimension $d$.\cite{Hatcher2009}

\subsection{Including point-group symmetry}\la{sec:pointgroup}



The inclusion of  point-group symmetry allows us to refine the notion of localizable bands -- into band representations and localizable topological bands. While the topological triviality of an $N$-band BR ($N{\geq}1$) is not difficult to show (and will be shown in \s{sec:analytictrivial}), the bulk of this letter addresses the converse question: is a topologically trivial $N$-band subspace with $G$ symmetry necessarily a BR of $G$? The existence of Kane-Mele topological insulators in $d{=}2,3$ (which necessarily have vanishing first Chern class $c_1{=}0$ due to time-reversal symmetry)\cite{Panati2007} demonstrates that the answer is negative for even $N$; $N$ here is even due to Kramers degeneracy in class AII. Our contribution is to prove the positive answer for $N{=}1$ and for any symmorphic magnetic space group $G$ in any spatial dimension $d$. By `symmorphic',  we mean $G$ is  a semidirect product of its translational subgroup $T{\subset}G$ with the quotient group $G/T$. This $N{=}1$ result may be viewed as a symmetry-refined analog of the no-go statement mentioned in \s{sec:trans}: not only are simple bands unable to realize nontrivial higher Chern classes, we find that they also cannot realize   localizable topological insulators in  symmorphic magnetic space groups.

\section{Statement of theorem}\la{sec:theorem}

Our main result is encapsulated in the following theorem, which applies for  any spatial dimension $d$.

\begin{theorem}\la{theorem}
For any simple band whose corresponding projection operator is analytic throughout $T^d$,  being a band representation of a symmorphic magnetic space group $G$ is equivalent to being topologically trivial in the category of complex, unit-rank vector bundles. The latter condition is known to be equivalent\cite{Hatcher2009} to a vanishing first Chern class.
\end{theorem}

Let us discuss the physical scenarios where the above-stated assumptions on the simple band are attained. The simple band may be  an energy eigenfunction of a $G$-symmetric Schr\"odinger Hamiltonian $H_0{=}{-}{\bigtriangleup}{+}V(\br)$ with energy eigenvalue $\var_{\bk}$ that is nondegenerate at all $\bk$; this nondegeneracy shall be referred to as a gap condition. To ensure that $e^{-i\bk\cdot \br} H_0\eikr$, at each $\bk$, is self-adjoint, analytic,\footnote{Precisely, $e^{-i\bk\cdot\br}H_0\eikr$ should be in the entire analytic family of type A, as defined in \ocite{Reed}.} and has a point spectrum, $V(\br)$ has to satisfy  certain  physically reasonable conditions,\footnote{Dimension-dependent conditions on $V$ are stated in \ocite{Reed}, Theorem XIII.99: $V{\in}L^p($unit cell$,d^d\br)$ for $p{=}2$ if $d{\le}3$, $p{>}2$ if $d{=}4$ and $p{=}d/2$ if $d{\ge}5$.}  e.g., for $d{\le}3$ it is sufficient that $V$ is square-integrable over the primitive unit cell. Alternatively, $H_0$ might be a tight-binding Hamiltonian whose matrix elements decay exponentially in real space, which ensures that the Fourier transform of $H_0$ is analytic throughout $T^d$.\cite{Cloizeaux1964} The analyticity and gap conditions ensure that the projection operator to the simple band is analytic at all $\bk$.\cite{Reed,Nenciu1991,Panati2007,Read2017}

\section{Preliminaries }\la{sec:preliminaries}




\subsection{Space groups}

For any magnetic space group $G$ that is symmorphic, there exists a point where each  $g{\in}G$  is the composition [denoted $(p|\bR)$] of a transformation $p$ that preserves said point, and a translation by a Bravais-lattice vector $\bR$; generally $p{=}p(g)$ and $\bR{=}\bR(g)$, but we shall omit the arguments. We employ $BL$ as a shorthand for the Bravais lattice (e.g., $\bR{\in}BL$) and $RL$ for the dual (reciprocal) lattice to $BL$.

The set of $p$ defines the point group $\calp$ of $G$, which is isomorphic to $G/T$.\cite{Evarestov} $p$ is specified by: (i) a $d$-by-$d$ real orthogonal matrix $\check{p}$ that acts on real space, as well as (ii) a $\Z_2$ index $s_p{=}{\pm}1$ which indicates whether or not $p$ involves a time reversal operation; $g{=}(p|\bR)$ then acts on $(d{+}1)$-dimensional spacetime as $\br{\rightarrow}g\circ \br{:}{=}\check p \br{+}\bR$ and $t{\rightarrow} s_p t$. In magnetic space groups without time-reversal symmetry, $s_p{=}1$ for all $p{\in}\calp$. The standard multiplication rule for magnetic space groups is
\e{(q|\bR')(p|\bR){=}(qp|\check{q}\bR+\bR') \iwith s_{qp}=s_{q}s_p,\la{mrule}}
where the presence of $\check{q}$  reflects the noncommutativity of translations and point-group operations. 

A notion that is useful to characterize Wannier functions is a Wyckoff position $\bvarpi$ of $G$; $\bvarpi$ is defined as a spatial coordinate in $\R^d$ with an associated symmetry group $G_{\varpi}{\subset}G$. $G_{\varpi}$, the site stabilizer, comprises all elements of $G$ that preserve $\bvarpi$, i.e., for any $g{\in}G_{\varpi}$, $g{\circ}\bvarpi{=}\bvarpi$. If $g{\circ}\bvarpi{-}\bvarpi$ is a BL vector for all $g{\in}G$ (equivalently, $G/G_{\varpi}{\cong}T$), then we say that $\bvarpi$ has unit multiplicity. 






\subsection{General representations of space groups}\la{general}

A simple band is spanned at each $\bk$ by a Bloch function $\psi_{\bk}$, whose phase is not uniquely defined. The projection $P(\bk){=}\ket{\psi_{\bk}}\bra{\psi_{\bk}}$ is periodic over $T^d$.   By the assumptions stated in the theorem, $P(\bk)$ is  analytic at all $\bk$, and  $\psi_{\bk}$ forms a general representation of $G$. By this, we mean there exists  a map from $g\in G$ to a unitary $\rho_g(\bk) {\in} U(1)$, such that 
\e{ \hat g \, \psi_{\bk}(\br) = \rho_g (\bk) \psi_{s_p\check p \bk}(\br). \la{repdef} }
Here, $\hat{g}$ is defined as a representation of $G$ that acts on functions  of real space as $\hat{g}f(\br){=} \overline{f(g^{\mo}{\circ} \br)}^{s_p}$, where $\bar{a}^{1}{:=}a$  and $\bar{a}^{{-}1}{:=}\bar{a}$ (the complex conjugate).\footnote{This action of $g$ on $f$ defines a regular representation\cite{Bacry1993}  which is known to be linear.} While \q{repdef} and the remaining proof is specific to Schr\"odinger wavefunctions, the proof is essentially unchanged if we replace $\psi_{\bk}(\br)$ by a finite-dimensional vector in a tight-binding basis of L\"owdin-orthogonalized orbitals.\cite{Lodwin1950} 

From the action of $\hat g$ on $\psi_{\bk}(\br)$, we deduce that $\rho_g$ may be factorized into translational and point-group components as
\e{\pdg{\rho}_{\sma{(p|\bR)}}{(\bk)}{=}\overline{{\rho}_{\sma{ (E|\bR)}}(\check p \bk)}^{s_p}\pdg{\rho}_{\sma{(p|\bze)}}(\bk){=}e^{\minus i s_p\check{p}\bk\cdot \bR}\pdg{\rho}_{\sma{(p|\bze)}}(\bk), \la{factorization}}
where $E$ denotes the identity element of the point group, and $\rho_{(E|\bR)}(\bk) {=}\mathrm{e}^{{-}i \bk \cdot \bR}$ describes the translational property of Bloch functions. That such a factorization exists reflects that $G$ is a semidirect product of its translational- and point-subgroups. A useful implication of \q{factorization} is that $\rho_{\sma{(p|\bR)}}(\bze){=}\rho_{\sma{(p|\bze)}}(\bze)$ is independent of $\bR$. Owing to 
\e{  \hat{h} \big( \hat{g}f(\br) \big) =\overline{f\big(\,g^{\mo}{\circ}( h^{\mo}{\circ} \br)\,\big)}^{s_{q}s_{p}}=  \overline{f\big((hg)^{\mo}{\circ} \br\big)}^{s_{qp}}}
for all $g{=}(p|\bR),h{=}(q|\bS) {\in} G$ and $hg$ defined through \q{mrule}, the representation $\hat{g}$ is linear (i.e., $\hat{h}\hat{g}{=} \widehat{hg}$), 
and therefore
\e{  \pdg{\rho}_{\sma{(q|\bS)}}\big(\pdg{s}_{p}\check{p} \bk\big) \,\overline{{\rho}_{\sma{(p|\bR)}}(\bk)}^{\pdg{s}_{q}} = \pdg{\rho}_{\sma{(q|\bS)(p|\bR)}}(\bk). \la{app:cocycle} }

\subsection{Localizable representations  of space groups}\la{sec:wannierrep}

Under a phase redefinition (or change in gauge) of $\psi_{\bk}$, $\psi_{\bk}$ and $\rho_g(\bk)$ transform as 
\e{{ \psi_{\bk} } \rightarrow \mathrm{e}^{i \phi(\bk)}{ \psi_{\bk} },\;\; \rho_{g}(\bk) \rightarrow e^{-i \phi(s_p\check p \bk)} \rho_{g}(\bk) e^{i s_p\phi(\bk)}. \la{eq:gaugetransform}}
Under such a gauge transformation, $\psi_{\bk}$ can be made analytic at $\bk$, for every $\bk{\in}T^d$; the existence of such  analytic local sections is guaranteed by the assumed analyticity of the projection $P(\bk)$.\footnote{This follows from analytic perturbation theory, e.g., see \ocite{Dubail2015} for the tight-binding $H_0$, and the Kato-Rellich theorem in \ocite{Reed} for the Schroedinger $H_0$.} Whether $\psi_{\bk}$ can be made analytic {throughout} and periodic over $T^d$ depends not just on the analyticity of $P(\bk)$, but also requires that there are no topological obstructions in the category of complex vector bundles.\cite{Nenciu1991,Panati2007} That is, if the simple band is topologically trivial, $\psi_{\bk}$ exists that is continuous and periodic over $T^d$; the continuity condition on $\psi_{\bk}$ can be further strengthened to analyticity throughout $T^d$.\footnote{A complex neighborhood of $T^d$ may be identified as a domain of holomorphy in $\C^d$ and therefore a Stein space. Solving the second Cousin problem over a Stein space is equivalent to proving the existence of a global analytic section for a topologically trivial line bundle; this has been carried out in \ocite{Hormander1989}. See also related discussions falling under the `Grauert-Oka principle'\cite{Oka1939,Grauert1958} in \ocite{Panati2007,Huckleberry2013}. In fact, the non-abelian second Cousin problem has also been solved using sheaf theory;\cite{Grauert1958,Gindikin1986,Henkin1998} this implies, for a topologically trivial band of rank $N{\geq}1$, that there exists analytic and periodic Bloch functions which span the $N$-dimensional vector space at each $\bk$.}



Henceforth it should be understood that any function of $\bk$ that is described as `periodic' (resp.\ `analytic') is periodic over $T^d$ (resp.\ analytic throughout $T^d$). Being both analytic and periodic  are necessary and sufficient\cite{Cloizeaux1964} conditions for the Fourier transform of $\psi_{\bk}$   
\e{ \pdg{w}_{\bR}(\br)  = \f{1}{\sqrt{{|T^d|}}} \int_{T^d} \mathrm{e}^{-i \bk \cdot \bR}  \psi_{\bk}(\br)  \mathrm d\bk \la{app:Wdef}}
to be exponentially localized; $|T^d|$ above denotes the volume of $T^d$. Such a localized wavepacket is referred to as a Wannier function; any band subspace which forms a general representation of $G$ and is unitarily equivalent to a set of Wannier functions is said to be a localizable representation of $G$.

\subsection{Band representations  of space groups}\la{sec:bandrep}

As motivated in the introduction, not all localizable representations of $G$ are BRs of $G$. For the purpose of proving our theorem, we may specialize the definition of BRs to simple bands with symmorphic $G$ symmetry: a  localizable representation of $G$ given by $\{w_{\bR}\}_{\sma{\bR\in BL}}$ [cf.\ \q{app:Wdef}] is a BR of $G$ with unit-multiplicity Wyckoff position $\bvarpi$, if  $w_{\bR}$ forms a representation of the site stabilizer $G_{\bvarpi{+}\bR}{\cong}\calp$, for any $\bR{\in}BL$. This may be viewed as a precise restatement of the local symmetry condition first formulated in the introduction. The general definition of BRs that is applicable to non-simple bands and nonsymmorphic space groups is provided in \app{app:equivdef}.

It is instructive to physically interpret\cite{Michel1999} $\bvarpi$ as the Wannier center:
\e{ \bar{\br} := \braket{ w_{\bze}| \br| w_{\bze} } = \f{1}{|T^d|} \int_{T^d} \bA(\bk) \mathrm d^dk, \la{app:pol} }
with $\br$ the position operator; the last equality utilizes a known relation between polarization and an integral of the Berry connection $\bA(\bk)$.\cite{King-Smith1993,Vanderbilt1993,Resta1994} In this interpretation, a Wannier function centered at $\bar{\br}{+}\bR$ forms a representation of the site stabilizer $G_{\bar \br+\bR}$.

The following lemma is useful to prove our theorem: a sufficient condition for band representability is that a general representation satisfies  
\e{\forall g =(p|\bS) \in G,\;\; \rho_{g}(\bk) \eq \pdg{\rho}_{\sma{(p|\bze)}}(\textbf{0}) \mathrm{e}^{-i s_p \check p \bk \cdot \bDelta_{{g}}}; \la{app:EBR}\\
 \bDelta_{{g}} :\eq g \circ \bvarpi {-} \bvarpi \in BL, \la{app:Brav} }
with $g$-independent $\bvarpi{\in} \R^d$. \qq{app:EBR}{app:Brav} shall be referred to as the canonical form of a BR.\cite{Zak1979} Especially, $\rho_g(\bk)$ depends on $\bS$ only through $\Delta_g$.

\noindent \emph{Proof of lemma.} \q{app:Brav} is the defining property for a unit-multiplicity Wyckoff position $\bvarpi$.  Any element in $G_{\varpi}$ has the form $p_{\bvarpi}{:}{=}(p|{-}\bDelta_{\sma{(p|\bze)}})$ with $p{\in}\calp$; this reflects an isomorphism with $\calp$. Combining \qq{app:EBR}{app:Brav} with \q{app:Wdef}, we derive a unitarily equivalent representation of $G$ on Wannier functions:
\e{ \forall g=(p|\bS)\in G,\as \hat g \ket{ w_{\bR} } = \pdg{\rho}_{\sma{(p|\bze)}}(\textbf{0}) \ket{ w_{\check p \bR + \bDelta_{{g}}} }. \la{repofG} } 
To interpret \q{repofG}, $\hat g$ has a two-fold effect (i) of translating the Wannier center $\bR{+}\bvarpi$  to ${g}{\circ}(\bR{+}\bvarpi){=}\check p \bR {+} \bDelta_{{g}} {+} \bvarpi$, and,  additionally, (ii) $\hat{g}$ may  transform the Wannier function around its own center, thus inducing the  phase factor $\rho_{\sma{(p|\bze)}}(\textbf{0})$. Let us demonstrate that \q{repofG} describes a BR of $G$. Restricting \q{repofG} to $\bR{=}\boldsymbol{0}$ and ${p}_{\bvarpi}{\in} G_{\varpi}$, we derive that $\hat p_{\bvarpi} \ket{w_{\bze}}{=}\rho_{\sma{(p|\bze)}}(\bze)\ket{w_{\bze}}$, i.e., $\rho_{\sma{(p|\bze)}}(\bze)$  is a representation of $G_{\varpi}$ that is restricted from $G$. One may {further verify} that $\ket{ w_{\bR} }{=}\widehat{(E|\bR)} \ket{w_{\bze}}$ [cf.\ \q{app:Wdef}] forms a representation of 
\e{ G_{\varpi+R}= (E|\bR) \,G_{\varpi}\, (E|\bR)^{\mo} \la{groupR}}
for any $\bR{\in} BL$, which proves the lemma.  


It is instructive to demonstrate that the simple band represented by \qq{app:EBR}{app:Brav} satisfies the conventional\cite{Zak1981,Evarestov1984,Bacry1988} definition of a BR: as a representation of $G$ that is induced from a representation of a site stabilizer $G_{\varpi}$ on a Wannier function, for some Wyckoff position $\bvarpi$. Indeed, were we to carry out this induction, we would expand the representation space of $w_{\bze}$ to include all $BL$-translates of $w_{\bze}$;\footnote{Generally, one constructs Wannier functions for all Wyckoff positions in the orbit $G{\circ}\bvarpi$; for unit-multiplicity Wyckoff positions, these Wannier functions are just the $BL$-translates of the single Wannier function.} these Wannier functions transform as
\e{ \forall (p|\bS){\in} G, \;\;\widehat{(p|\bS)} \ket{w_{\bR}} \eq \widehat{(E|\check p \bR{+}\bDelta_{\sma{(p|\bS)}})}\hat{p}_{\bvarpi}\ket{w_{\bze}},}
from which we recover \q{repofG}. This proves that \q{repofG} is a BR as conventionally defined.

\section{Band representations are trivial}\la{sec:analytictrivial}

To recapitulate, a simple BR is a Wannier  representation of $G$, such that each Wannier function  represents its site stabilizer. By Fourier transformation [the inverse of \q{app:Wdef}], we obtain a Bloch function that is analytic and periodic.\cite{Cloizeaux1964} More generally, an $N$-band BR ($N{\geq}1$) is analytically trivial, i.e., there exist $N$ Bloch functions which are periodic and analytic, and span the $N$-band BR at each $\bk$. Being analytically trivial is generally a stronger condition than being topologically trivial.

\section{A simple topologically trivial band is a band representation}\la{sec:forwardarrow}
 
For a simple, topologically trivial band, we have argued that $\psi_{\bk}$ can be made analytic and periodic; henceforth these properties are assumed for $\psi_{\bk}$; it follows from \q{repdef} that both properties are likewise satisfied by $\rho_g(\bk)$. Under this assumption, there remains a freedom  to perform gauge transformations [cf.\ \q{eq:gaugetransform}] with $e^{i\phi(\bk)}$ that is analytic and periodic. Exploiting this freedom, we would show $\rho_g(\bk)$ may be simplified to the canonical form [cf.\ \qq{app:EBR}{app:Brav}] for all $g{\in}G$; according to the lemma in \s{sec:bandrep}, this would prove the desired result.

It is not difficult to see that the canonical form applies to the translational subgroup of $G$: $\rho_{(E|\bR)}(\bk){=}e^{{-}i\bk\cdot \bR}$, as derived in \qq{repdef}{factorization}. Owing to the simple factorization of  $\rho_g$  [cf.\ \q{factorization}] for symmorphic space groups, what remains is to prove the canonical form for the point-preserving elements of $G$: $\{(p|{\bze})|p{\in}\calp\}$. To simplify notation in the rest of this section, we shorten $(p|{\bze})$ to $p$, e.g., $\rho_{\sma{(p|{\bze})}}{\equiv}\rho_{\sma{p}}$. We split the proof into three steps, which are to be proven for all $p{\in}\calp$.  
\begin{enumerate}
\item A general form for $\rho_p(\bk)$ is 
\e{ \rho_p(\bk)=\mathrm{e}^{-i s_p\check p \bk \cdot \bDelta_{{p}} + i \alpha_p(\bk)}, \as\text{with}\;\;\bDelta_{{p}}{\in}BL \la{app:genform} }
and $\alpha_p(\bk)$ a real, analytic, periodic function. 



\item There exists $\bvarpi$ such that $\bDelta_{{p}}{\in}BL$ in \q{app:genform} satisfies \q{app:Brav}. 

\item By applying a further gauge transformation, the periodic component of the phase of $\rho_p$ [cf.\ \q{app:genform}] may be made independent of $\bk$: $\alpha_p(\bk) {\to} \alpha_p(\textbf{0})$.
\end{enumerate}

\noindent 1.-3. then imply \qq{app:EBR}{app:Brav} with the identification $\mathrm{e}^{i \alpha_p(\textbf{0})} {=} \rho_{p}(\textbf{0})$ for all $p{\in} \calp$. 


\subsection{Proof of 1.}





$\rho_p(\bk)$ is a map from the $d$-torus to $U(1)$, and the homotopy classes of such maps are classified by $d$ integers;\footnote{This follows because $[T^d,U(1)]$ are in one-to-one correspondence with cohomology classes in $H^1(T^d;\Z)$. Applying the Universal Coefficient Theorem, $H^1(T^d;\Z){\cong}\Z^d$. } these integers may be identified as winding numbers (denoted $n_1,{\ldots}n_d{\in}\Z$) of the $U(1)$ phase over $d$ independent primitive vectors $(\bG_1,{\ldots},\bG_d){\subset} RL$. Equivalently stated, if we define  $\theta_p(\bk)$ as the phase of $\rho_p(\bk)$ such that $\theta_p(\bk)$ is analytic throughout $\R^d$; then $\theta_p(\bk{+}\bG_j){=}\theta_p(\bk){+}2\pi n_j$. Without loss of generality, we may decompose $\theta_p$ into periodic ($\alpha_p$) and nonperiodic components as  
\e{\theta_p(\bk)=\alpha_p(\bk)+\bk\cdot \sum_{i=1}^dn_i\bS_i ; \as \bS_i\cdot \bG_j=2\pi \delta_{ij},} \\
where $\bS_i{\in} BL$ are primitive Bravais lattice vectors dual to $\bG_j$. Since ${\sum}_i n_i \bS_i{\in}BL$ and the Bravais lattice has the symmetry of the point group $\calp$, there is no loss in generality in expressing ${\sum}_in_i\bS_i{=}{-}s_p\check p^{-1} \bDelta_{{p}}$ with $\bDelta_{{p}}{\in}BL$ and $p{\in}\calp$. This completes the proof of 1.




\subsection{Proof of 2.}\la{proof2}

\q{app:cocycle} constrains the phases of $\rho_p(\bk)$ as:
\e{ &s_q\alpha_p(\bk) + \alpha_q(s_p\check p \bk) - \alpha_{qp}(\bk) - 2\pi n(q,p)   \la{alphas} \\
\eq s_{pq}\big(\,\check p \bk \cdot \bDelta_{{p}}+\check q \check p \bk \cdot \bDelta_{{q}} -\check q \check p \bk \cdot \bDelta_{{qp}}\,\big)  \la{app:comult} }
for all $p,q {\in} \calp$; $n(q,p) {\in} \Z$ is introduced to account for the $2\pi$-ambiguity of the phase. That $n(q,p)$ is independent of $\bk$ follows from the analyticity of $\alpha_{p}(\bk)$ and the integer-valuedness of $n(q,p)$. 

Let us demonstrate that the left- [\q{alphas}] and right-hand-sides [\q{app:comult}] of the above equality  vanishes separately. Since $\alpha_p(\bk)$ is analytic, we may apply the gradient $\nabla_{\bk}$ to \qq{alphas}{app:comult}:
\e{ s_q \nabla_{\bk} \alpha_p(\bk) + s_p\check p^{-1} \nabla_{s_p\check p \bk} \alpha_q(s_p\check p \bk) -\nabla_{\bk} \alpha_{qp}(\bk) \la{firstfirst} \\
=   s_{qp}\check p^{-1} \big(\,\bDelta_{{p}}+  \check q^{-1} \bDelta_{{q}} - \check q^{-1} \bDelta_{{qp}}\,\big). \la{app:grmult} }
The periodicity of $\alpha_p$ and all  terms in \qq{firstfirst}{app:grmult} allows for a Fourier analysis; this demonstrates that $\nabla_{\bk} \alpha_p(\bk)$ does not contain a constant-in-$\bk$ term, hence the bracketed terms in \q{app:grmult} vanish by linear independence.  Summing these bracketed terms over all $q {\in} \calp$ and dividing by the order ($|\calp|$) of $\calp$, we derive \q{app:Brav} with
\e{ \bvarpi = \f{1}{|\calp|} \sum_{q \in \calp} \check q^{-1} \bDelta_{{q}}. \la{app:wcenter}}





\subsection{Proof of 3.}\la{proof3}


Having determined that \q{alphas} vanishes for all $p,q {\in} \calp$, we multiply it by $s_q$, sum over all $q {\in} \calp$ and divide by $|\calp|$ to obtain
\e{ \alpha_{p}(\bk) = s_p\Phi(\bk) - \Phi(s_p \check p \bk) + \f{2\pi}{|\calp|} \sum_{q \in \calp} s_q n(q,p). \la{app:alpha}}
where $\Phi(\bk) {:}{=}  {\sum}_{q \in \calp}s_q \alpha_{q}(\bk)/{|\calp|}$. Since $\Phi(\bk)$ is independent of $p$, the $\bk$-dependent terms on the right-hand-side of \q{app:alpha} can be removed by a gauge transformation [\q{eq:gaugetransform}]; we may view this as a transformation between two homotopically equivalent representations.\footnote{For any two representations $\big( \rho_p^0(\bk),\rho_p^1(\bk) \big)$ of $G$ that are related by a gauge transformation with periodic $\phi(\bk)$, there exists a continuous interpolation $\rho_p^s(\bk)$ ($s\in [0,1]$) that is itself analytic and periodic, and a representation of $G$ throughout the interpolation.} This completes the proof of the theorem. We provide a group-cohomological perspective of $n(q,p)$ as a two-cocyle in \app{app:remarkcohomological}.

\section{Generalization to solids with spin-orbit coupling}\la{sec:generalizeSOC}

Our theorem is generalizable to spin-orbit-coupled solids with broken time-reversal symmetry, where the absence of Kramers degeneracy allows for simple bands. The statement of the theorem for spin systems is nearly identical to the spinless case, except $G$ is now identified with a double\cite{MelvinLax1974} symmorphic magnetic space group. The proof of the theorem with spin is essentially identical, except $\psi_{\bk}$ should be replaced by a spinor wavefunction, and  $\calp{\cong}G_{\varpi}$ now includes a non-identity element\cite{Tinkham2003} corresponding to a $2\pi$ rotation.


\section{Applications of the theorem}\la{sec:applications}

Our theorem   (including the generalization to half-integer-spin representations in \s{sec:generalizeSOC}) has two types of applications: the first  rules out simple localizable topological bands, which implies a minimal dimension for the tight-binding Hamiltonian of localizable topological insulators [\s{sec:impossible}], and the second guarantees simple Chern insulators [\s{sec:chern}]. An application to solids with spacetime-inversion symmetry is highlighted in \s{sec:generalizeinversion}, which illustrates a Stiefel-Whitney obstruction that occurs only for real vector bundles.

\subsection{A simple band cannot be localizable topological}\la{sec:impossible}

The following discussion applies in any spatial dimension $d$, and for  $G$ that is symmorphic. A corollary of Theorem \ref{theorem} states that a simple band cannot be localizable topological, i.e., a localizable topological band minimally has rank two. There are two classes of localizable topological bands distinguished by their stability upon summation\footnote{Precisely, we mean a Whitney sum of two vector bundles, where at each $\bk$ the two vector spaces are directly summed.} with a BR of $G$: (i) a stable localizable topological band remains localizable topological upon summation (as exemplified by the Kane-Mele phase),\cite{kane2005A} while (ii) a fragile localizable topological band becomes band-representable upon summation (as exemplified in \ocite{Po}). 

\subsubsection{Minimal rank for the tight-binding Hamiltonian of a localizable topological insulator}\la{sec:impossible2}

Restricting our discussion to $d{\leq}3$, we deduce that the minimal rank of a tight-binding Hamiltonian for a fragile localizable topological insulator is three, and that for a stable localizable topological insulator is four. By `rank' of a tight-binding Hamiltonian, we mean the dimension of the tight-binding Hilbert space as restricted to a wavevector, or to one real-space unit cell.
Indeed, a $G$-symmetric tight-binding Hilbert space is also a BR of $G$;\cite{Holler2017}  this BR must be split to attain a localizable topological band (fragile or topological). Minimally, the tight-binding BR splits into two band subspaces, which we may denote as filled and empty. For a localizable topological filled band, it must be that the empty band is also localizable (i.e., $c_1{=}0$); this follows because the tight-binding BR must be localizable, and the first Chern numbers are stable invariants.\cite{kitaev_periodictable} 

If the filled band is stable localizable topological, then so must the empty band (by the definitions of fragile and stable given above), and therefore the combined minimal rank is four. This minimal rank is saturated by the Kane-Mele model\cite{kane2005A} of the $\Z_2$ topological insulator, as well as the Fu model\cite{Fu2011} for a rotationally-symmetric topological crystalline insulator. 
On the other hand, if the filled band is fragile localizable topological, the empty band must be band-representable and its rank is not constrained by our theorem; then the combined minimal rank is three. This minimal rank is saturated by a Kagome model that is detailed below [cf.\ \s{loctop}]. These rank constraints on the tight-binding Hamiltonian are supported empirically by all localizable topological insulators that we know,\cite{Fu2011,Liu2014,Alexandradinata2014g,Alexandradinata2014c,Shiozaki2014,Fang2015,Shiozaki2015,Hourglass2016,Shiozaki2016,Shiozaki2017,kane2005A,Kane2005b,QSHE_bernevig,HgTe_bernevig,QSHE_Rahul,fu2006,fukanemele_3DTI,Inversion_Fu,moore2007,Rahul_3DTI,Classification_Chiu,AZ_mirror,Bradlyn2017,Cano2017} and guides future work in modelling yet-unknown phases.





\subsection{Program for mass identification of topological materials}\la{sec:chern}

 In combination with the theory of elementary band representations (EBRs),\cite{Zak1979,Zak1981,Bacry1988} our theorem  guides the design and identification of Chern- and fragile localizable topological insulators. EBRs are the basic building blocks of space-group representations, and serve a role analogous to irreducible representations or finite groups.\cite{Bacry1993} An EBR is defined as a BR of $G$ which cannot be split  into multiple fewer-band subspaces that are all BRs of $G$; if an EBR of $G$ were splittable, then at least one of the fewer-band subspaces cannot be a BR of $G$.\cite{Holler2017,Bradlyn2017,Cano2017,Po} Most EBRs satisfy two properties: (i) its Wyckoff position $\bvarpi$ is maximal, and (ii) the representation of $G_{\varpi}$ is irreducible;\cite{Bacry1988,Bradlyn2017,Cano2017} the exceptions to (i{-}ii) are tabulated in the given references. 

Whether an EBR is splittable is determined by compatibility relations in combining symmetry representations of little groups over the Brillouin torus;\cite{Bradlyn2017} these combinations are tabulated in the \href{http://www.cryst.ehu.es/}{Bilbao Crystallographic Server},\cite{bilbao} for space groups with and without time-reversal symmetry and $d{=}3$. In more detail, under the link `BANDREP', all  splittable EBRs of a specified space group are labelled as `Decomposable'; under the link `Decomposable', all possible connected subspaces (labelled `branch 1', `branch 2', {\ldots}) of a splittable EBR are specified by their irreducible representations of little groups at high-symmetry quasimomenta. By a `connected' $N{\geq}1$-band, we mean that symmetry-enforced band touchings prevent the band from being separated energetically into two or more components; by definition, a simple band is always connected. In symmorphic space groups, a connected subspace may be identified as simple if the representations of all little groups are one-dimensional. If one or more of these connected subspaces are simple, our theorem becomes useful in identifying Chern insulators [as discussed in \s{chern}] and fragile localizable topological insulators [\s{loctop}]. 





 \subsubsection{Identifying Chern insulators}\la{chern}
 
If an EBR were splittable into a simple band which is band-unrepresentable, then our theorem guarantees that it has a nonzero first Chern class (in short, it is a Chern band). For lower-symmetry space groups, it is not uncommon to find rank-$s$ EBRs that  split into $s{\geq}1$ simple bands; due to the net topological triviality of all $s$ bands, it follows that at least two of them must be Chern bands. Thus if $s{=}2$, and if the Fermi level separates the two simple bands, a Chern insulator is guaranteed. We have exemplified such EBRs on a 2D checkerboard lattice ($s{=}2$, wallpaper group $P4$), and a 2D honeycomb lattice ($s{=}2$, $P6$) in \ocite{Holler2017}.

For $s{\geq}3$ simple bands, further work is required to determine which of the $s$ simple bands are Chern bands, e.g., by a symmetry-representation\cite{Hughes2011a,Turner2012a,Fang2012,Hourglass2016,Song2017,Watanabe2017a} or a Zak-phase analysis\cite{Holler2017}. For illustration, we consider a three-band EBR comprising $s$-orbitals on a Kagome lattice (wallpaper group $P6$). In a tight-binding model with complex nearest-neighbor hoppings (e.g., $t{=}i$), the EBR splits into three components with first Chern numbers: $C_1{=}0, {+}1, {-}1$ respectively;\cite{Holler2017} a Chern insulator is attained if the Fermi level separates two subspaces with nontrivial Chern numbers.

After picking a candidate EBR of a group $G$, it is straightforward to design a tight-binding model of the Chern insulator; the tight-binding basis vectors would correspond directly to this EBR.\cite{Holler2017} By varying the $G$-symmetric tight-binding matrix elements, we may explore all possible splittings of this EBR, one of which would give Chern bands. Finally, to find a naturally-occurring Chern insulator, we propose to search for $G$-symmetric materials with our candidate EBR in the vicinity of the Fermi level.


\subsubsection{Identifying fragile localizable topological insulators}\la{loctop}

Suppose a rank-$s$ EBR splits into two orthogonal subspaces ($S_1,S_2$), each of which is not necessarily connected. Then if $S_1$ is band-representable, $S_2$ must be fragile localizable topological. If $S_1$ comprises of simple band(s) with vanishing first Chern number(s), then our theorem guarantees that $S_1$ is band-representable. An illustration is provided by splitting the same Kagome EBR just described in   \s{chern}: If $S_1$ is the simple band with $C_1{=}0$, then $S_2$ is composed of two simple bands with $C_1{=}{+}1,{-}1$, and must therefore be fragile localizable topological. This fragile topology is manifested by a nontrivial winding of the eigenvalues of the holonomy matrix (i.e., Wilson loop).\footnote{It is characterized by a unit relative winding number -- a topological invariant first discovered in \ocite{Alexandradinata2014c}. This invariant has also been used to characterize other fragile phases.\cite{Cano2017a}} 

Our theorem has greatest utility in magnetic space groups which guarantee the triviality of the first Chern class (e.g., time-reversal symmetry, or $(d-1)$ orthogonal mirror symmetries in a $d$-dimensional cubic lattice). For such groups, $S_1$ is guaranteed to be band-representable by our theorem. As an application, let us search in Bilbao for the following EBR of space group 183 with time-reversal symmetry: this EBR is labelled by the Wyckoff position $\bvarpi=3c$, and a trivial on-site symmetry representation $A_1$ of the on-site stabilizer $G_{\varpi}=C_{2v}$. This rank-three EBR is splittable into a simple band (`branch 1') and a pair of bands (`branch 2'); the latter must be fragile localizable topological.

\subsection{Application to bands with spacetime-inversion symmetry}\la{sec:generalizeinversion}

We remark on a class of bands that arise in solids with negligible spin-orbit coupling, and the symmetry of spacetime-inversion: $(\br,t){\rightarrow} ({-}\br,{-}t)$. This symmetry ensures that $h(\bk)$, the tight-binding Hamiltonian, can be made real at each $\bk$, as explained in \app{spacetime}. The real eigenvectors of $h(\bk)$ (corresponding to the filled bands) define a real vector bundle which is guaranteed to have vanishing first Chern class by the spacetime-inversion symmetry. Since a real vector bundle (or real band) can be embedded in a complex one (in analogy with how real numbers can be embedded in complex numbers),\cite{DeNittis2014} our theorem also applies to real bands of symmorphic magnetic space group $G$; in fact, the theorem directly implies that any real simple band with $G$ symmetry is necessarily a BR of $G$. 

Despite being trivial when viewed as a complex unit-rank (i.e., line) bundle, a real simple band  may nevertheless be nontrivial in the category of real line bundles. Triviality in the latter category is equivalent\cite{Hatcher2009} to the vanishing of the first Stiefel-Whitney characteristic class, or equivalently to the existence of a real section (a real eigenvector $\ket{u_{\bk}}$ of $h(\bk)$ that is continuous, periodic and nonvanishing over $T^d$). The latter implies that  the Berry connection $i\braket{u_{\bk}|{\nabk} u_{\bk}}$ (being necessarily real) must vanish at all $\bk$, and hence the  Berry-Zak phase vanishes for the holonomy over all noncontractible Brillouin-zone loops. Utilizing the geometric theory of polarization,\cite{King-Smith1993} we deduce that the corresponding Wannier center lies at the spatial origin, modulo translations by a Bravais-lattice vector.

An obstructed simple real band is exemplified by either of the two energy bands of the Hamiltonian $h(k){=}\cos(k)\sz{+}\sin(k)\sx$, which is real at all $k{\in}T^1$ ($\sx, \sz$ are Pauli matrices). A real eigenvector of $h(k)$ that is continuous at each $k$ must satisfy anti-periodic boundary conditions over $T^1$, i.e., the Berry-Zak phase is $\pi$. An analytic and periodic eigenvector can be attained by multiplying the real eigenfunctions of $h(k)$ by a complex phase factor $e^{ik/2}$ [cf.\ \q{eq:gaugetransform}]; the corresponding Wannier function is necessarily displaced by half a lattice period (modulo lattice translations) from the origin.

\section{Discussion and outlook}\la{sec:discussion}

The question of the minimal rank of a band subspace is increasingly enriched by the interplay between band topology and crystallographic symmetry.\cite{Michel1999,Parameswaran2013,Po2016,Watanabe2017a} A space-group-symmetric band subspace may be topologically trivial, yet its rank exceeds unity if there exists symmetry-enforced band-touching points in $T^d$. These touchings are associated to higher-than-one dimensional irreducible representations of the little group at high-symmetry $\bk{\in}T^d$,\cite{Bouckaert1936,Herring1937} or to the nontrivial monodromy of nonsymmorphic symmetry representations.\cite{Michel1999,Holler2017,Parameswaran2013,Po2016,Watanabe2016} This local-in-$\bk$ perspective of symmetry representations contrasts with a global perspective that is required to understand topological band insulators. A case in point are insulators with a nontrivial higher-than-one Chern class; their rank must exceed unity,\cite{Hatcher2009} independent of the presence of crystallographic symmetry. 


In this work, we present a rank constraint that relies saliently on both band topology and crystallographic point-group symmetry: the rank of a localizable topological band  must exceed unity. To recapitulate, we have defined a localizable topological band as having an obstruction to locally-symmetric Wannier functions, i.e., it is a localizable band that is not a band representation (BR). This local symmetry condition has been defined in our introduction, and a more elaborate definition may be found in \app{app:equivdef}. Our rank constraint is fundamentally different from those imposed by symmetry-enforced touching points in $\bk$-space, e.g., a fragile localizable topological band may be a sum of simple bands with no such touching points, as exemplified by the Kagome model in \s{loctop}.   

An alternative restatement of this rank constraint is that the phase diagram of unit-rank band subspaces (simple bands) is comparatively uncomplicated, as illustrated in \fig{fig:pie}(a-b): either a simple band is unlocalizable (with a nontrivial first Chern class) or otherwise it is a BR.  Our rank constraint is the corollary of a theorem that states: for simple bands, being a band representation is equivalent to being topologically trivial in the category of complex vector bundles. It should be emphasized that our theorem applies generally to space-group-symmetric band systems -- independent of the statistics of particles that fill the band. In particular, the applications extend to bosonic band systems such as photonic crystals,\cite{Lu2014,Lu2016} phonon bands,\cite{Susstrunk2015,Susstrunk2016} and linear circuit lattices.\cite{Ningyuan2015,Albert2015,Lee2018}


Finally, we remark on our definition of a BR -- as unitarily equivalent to locally-symmetric Wannier functions. A priori, this definition is not obviously equivalent to the conventional definition\cite{Zak1979,Zak1981,Evarestov1984,Bacry1988} of BRs as induced representations of space groups; this equivalence is proven in \app{app:equivdef}. Arguably, our unconventional definition more directly and intuitively pinpoints the difference between BRs and localizable topological bands.





\begin{acknowledgments}
We are  grateful to Nicholas Read for explaining the Oka-Grauert principle, which stimulated a generalization of our theorem to arbitrary spatial dimensions. Ken Shiozaki and Nicholas Read educated us on the Hopf problem and the Stiefel-Whitney classes. Wang Zhijun and Barry Bradlyn taught us how to use the Bilbao tables on elementary band representations. Charles Zhaoxi Xiong, Ken Shiozaki and Michael Freedman clarified a homotopy nicety. We thank Alexey Soluyanov, Nicholas Read and Wang Chong for a critical reading of the manuscript. The authors further acknowledge stimulating discussions with Meng Cheng, Andrei Bernevig, Jennifer Cano, Ashvin Vishwanath, Haruki Watanabe and Adrian Po.

We acknowledge support by the Yale Postdoctoral Prize Fellowship (AA) and NSF grant No. DMR-1408916 (JH).
\end{acknowledgments}

\appendix

\section{Equivalence of definitions of band representation}\la{app:equivdef}

Two equivalent definitions for band representations (BRs) of space groups exist: one from the perspective of inducing a representation of site stabilizer, and another from the dual perspective of restricting a localizable representation of a space group to a site stabilizer. In this appendix we prove their equivalence.




\subsection{General definition of localizable representations}\la{app:gendeflocal}

Let us first define a rank-$(N{\geq} 1)$ localizable representation of a magnetic space group $G$  (symmorphic or nonsymmorphic). \begin{definition}\la{def:localrep}
\normalfont A \textit{localizable representation of $G$} is an infinite-dimensional linear representation of  $G$ on Wannier functions $\{ \ket{w_{n,\bR}^{\alpha_n}} \}_{\alpha_n,n,\bR}$ where $\bR{\in}BL$, $n$ labels distinct Wannier centers $\bvarpi_n$ in one unit cell, and $\alpha_n$ distinguishes Wannier functions centered on the same coordinate $\bvarpi_n$. 
\end{definition}
\noindent Here and throughout this paper, Wannier functions are defined to be exponentially-localized  Fourier transforms of Bloch functions:
\e{ w_{n,\bR}^{\alpha_n}(\br)  = \sum_{m,\beta_m}\int_{T^d} \f{\mathrm{e}^{-i \bk \cdot \bR}}{\sqrt{{|T^d|}}}   [U(\bk)]_{m,n}^{\beta_m,\alpha_n} \psi_{m,\bk}^{\beta_m}(\br)  \mathrm d\bk, \la{app:Wdef2}}
The Bloch functions $\{\psi_{m,\bk}^{\beta_m}\}_{m,\beta_m}$  are assumed to be analytic and periodic, and they orthonormally span the $N$-dimensional vector space at each $\bk$.   As explained in the main text, such Bloch functions  exist  if and only if the $(N{\geq} 1)$-band subspace is topologically trivial as a complex vector bundle.\cite{Grauert1958,Gindikin1986,Henkin1998} $U(\bk)$, when viewed as a matrix with row index $(m,\beta_m)$ and column index $(n,\alpha_n)$, is unitary, periodic and analytic. The presence of this matrix in \q{app:Wdef2} reflects a choice of basis (or `gauge freedom') for the Wannier functions.\cite{Marzari1997} It follows  from \q{app:Wdef2} all Wannier functions labeled by the same $n$ and $\alpha_n$ are related by Bravais-lattice translations: 
\e{ \ket{w_{n,\bR}^{\alpha_n}}= \widehat{(E|\bR)} \ket{w_{n,\bze}^{\alpha_n}}. \la{mayaswell1}}
This is a convenient choice of basis that exploits the discrete translational symmetry. 

In Definition \ref{def:localrep}, we have organized Wannier functions according to their Wannier centers, which are  defined as expectation values of the position operator [cf.\ \q{app:pol} for a simple band]:
\e{ \bvarpi_n+\bR=\bra{w_{n,\bR}^{\alpha_n}}\br \ket{w_{n,\bR}^{\alpha_n}},\la{defwcenter}}
The number of Wannier functions centered on $\bvarpi_n$ might vary with $n$; the total number of Wannier functions centered within one unit cell is $N$.




It is useful to define a special type of localizable representation:
\begin{definition}\la{define:locsingle}
\normalfont A \textit{localizable representation of $G$ with a single Wyckoff position $\bvarpi$} is a localizable representation of  $G$ where all Wannier centers are related to $\bvarpi$ by symmetry.   
\end{definition}
\noindent We may take $\bvarpi{:}{=}\bvarpi_1$, and define the site stabilizer $G_{\varpi}$ is the subgroup of $G$ that preserves $\bvarpi$. Definition \ref{define:locsingle} implies, for each $n=1,\ldots,M$, there exists a representative element $g_n {\in} \calp/G_{\varpi}$ such that $g_n{\circ} \bvarpi{=}\bvarpi_n$, with $g_1$ being the identity element. $M{=} |\calp|/|G_{\varpi}|$ is defined as the multiplicity of the Wyckoff position $\bvarpi$. It follows from  \q{defwcenter} that $\hat g_n\ket{w_{1,\bze}^{\alpha_1}}$ is centered at $\bvarpi_n$, hence the space orthonormally spanned by $\{\hat g_n\ket{w_{1,\bze}^{\alpha_1}}\}_{\alpha_1}$ is a subspace of the space orthonormally spanned by $\{\ket{w_{n,\bze}^{\alpha_n}}\}_{\alpha_n}$; conversely, we may demonstrate that the  space spanned by $\{\hat g^{\mo}_n\ket{w_{n,\bze}^{\alpha_n}}\}_{\alpha_n}$ is a subspace of the space spanned by $\{\ket{w_{1,\bze}^{\alpha_1}}\}_{\alpha_1}$. Clearly then $\hat g_n$ is an isomorphism between two vector spaces, and we may as well define
\e{\ket{ w_{n,0}^{\alpha} } =\hat{g}_n\ket{ w_{1,0}^{\alpha} },   \la{mayaswell2} }
for all $n,\alpha$; presently, we may drop the subscript of $\alpha_n$. 


We remark that any localizable representation of $G$ is a direct sum of single-Wyckoff localizable representations (possibly with distinct Wyckoff positions).

\subsection{Conventional definition of band representations}

A BR is a special type of localizable representation.\cite{Zak1979,Zak1981}
\begin{definition}\la{BR1}
\normalfont A \textit{band representation of $G$ with Wyckoff position $\bvarpi$} is the induced representation of  $G_{\varpi}\subset G$. A direct sum of band representations (possibly with distinct Wyckoff positions) is also referred to as a band representation.
\end{definition}
\noindent In more detail, inducing a finite-dimensional representation $V$ of $G_{\varpi}$ on Wannier functions $\{ \ket{ w_{1,0}^{\alpha} } \}_{\alpha=1}^{\mathrm{dim} V}$ gives an infinite-dimensional representation of $G$ on Wannier functions $\{ \ket{ w_{n,\bR}^{\alpha} } \}_{\alpha,n,\bR}$ where $\bR{\in}BL$, $\alpha{=}1,{\ldots},\mathrm{dim} V$;  $n{=}1,{\ldots},M$ with $M$ the multiplicity as defined above. This induced representation is a localizable representation of $G$ with a single Wyckoff position $\bvarpi$, for which Wannier functions can be chosen to satisfy \qq{mayaswell1}{mayaswell2}. The induced representation on such Wannier functions is, for all $g{=}(p|\bt) {\in} G$:\cite{Evarestov1984,Holler2017}
\e{ \hat g \ket{ w_{n,\bR}^{\alpha} } &= \sum_{\beta=1}^{\dim V}  [ V( p_{n'}^{\mo} p p_n) ]_{\beta\alpha} \ket{ w_{{n'},\check p \bR+\Delta_{g,n',n}}^{\beta} } \lin
\Delta_{g,n',n} &= g \circ \bvarpi_n - \bvarpi_{n'} \in BL; \la{app:Brav2} }
where the action of $g$ on a vector $\br$ is $g{\circ} \br{=} \check p \br {+} \bt$; if $g$ is a nonsymmorphic element, then $\bt$ is not a Bravais-lattice vector.

To motivate the form of \q{app:Brav2}, the action of $\hat{g}$ on Wannier functions may be separated into two effects: (i) a translation of the Wannier center from 
\e{\bvarpi_n+ \bR \rightarrow g\circ (\bvarpi_n+\bR) = \bvarpi_{n'}+\check p \bR+\Delta_{g,n',n},}
as well as (ii) a local transformation of Wannier functions sharing the same center, as effected by the matrix $V(p_{n'}^{\mo} p p_n)$. To motivate the argument of $V$, 
observe that $p_{n'}^{\mo} p p_n$ is the origin-preserving component of $g_{n'}^{-1} (E|{-}\Delta_{g,n',n}) g g_n $, which consecutively maps $\bvarpi {\rightarrow} \bvarpi_n {\rightarrow} g\circ \bvarpi_n {\rightarrow} \bvarpi_{n'}{\rightarrow} \bvarpi$. This implies that $g_{n'}^{-1} (E|{-}\Delta_{g,n',n}) g g_n $ is an element in the site-stabilizer $G_{\varpi}$, and $V$ a representation of $G_{\varpi}$. Note further that for all $g=(p|\bt)\in G_{\varpi}$, the set $(p|\bze)$  forms an isomorphic point group related to $G_{\varpi}$ by conjugation, hence we may as well label $V(g)$ by $V(p)$, as we have done in \q{app:Brav2}.


Corresponding to \q{app:Brav2} is the following representation of $G$ on $N {=} M {\times} \mathrm{dim} V$ Bloch functions:\cite{Evarestov1984,Holler2017}
\e{ [ \rho_g(\bk) ]^{\beta,\alpha}_{n',n} = \mathrm{e}^{-i s_p \check p \bk \cdot \Delta_{g,n',n}} [ V( p_{n'}^{\mo} p p_n) ]_{\beta\alpha}; \la{app:blochsymm} }
this may be derived by combining \q{app:Brav2} and \q{app:Wdef2}, with the choice $[U(\bk)]_{m,n}^{\beta,\alpha} = \delta_{m,n} \delta_{\beta,\alpha}$. $\rho_g(\bk)$, when viewed as a matrix with row index $(\beta,n')$ and column index $(\alpha,n)$, is unitary. \qq{app:Brav2}{app:blochsymm} are the generalizations of \qq{app:EBR}{repofG} to  $M\geq 1$ and $\mathrm{dim} V\geq 1$. 

\subsection{Equivalent definition of band representations}

We now provide an equivalent definition of a band representation -- by restricting representations of space groups to site stabilizers.
\begin{definition}\la{BR2}
\normalfont A \textit{band representation of $G$ with Wyckoff position $\bvarpi$} is a localizable representation of $G$ with $\bvarpi$, such that for all $n$ and all $\bR {\in} BL$, $\{ \ket{ w_{n,\bR}^{\alpha} } \}_{\alpha}$ forms a representation of the site stabilizer $G_{\varpi_n+R}$.    
\end{definition}
\noindent Definition \ref{BR2} means that  for all $n{=}1,{\ldots},M$ (the multiplicity) and all $\bR {\in}$BL, there exists a finite-dimensional unitary representation $X_{n,\bR}(p)$ of $G_{\varpi_n+R}$, i.e., for all $g{=}(p|\bS) {\in} G_{\varpi_n+R}$:
\e{ \hat g \ket{ w_{n,\bR}^{\alpha} } = \sum_{\beta=1}^{\dim X_{n,\bR}} \big[ X_{n,\bR}(p) \big]_{\beta,\alpha} \ket{ w_{n,\bR}^{\beta} }. \la{BRW2} }
In comparison, for any localizable representation of $G$, the full representation space $\{ \ket{ w_{n,\bR}^{\alpha} } \}_{\alpha,n,\bR}$ forms a representation of any subgroup of $G$. Since $g_n$ and $(E|\bR)$ act bijectively on Wannier centers [cf.\ \q{defwcenter}], they also induce an isomorphism of vector spaces spanned by Wannier functions (whose centers are related through \q{defwcenter}); an analogous demonstration has been provided in \s{app:gendeflocal}. Consequently, we may as well define \q{mayaswell1} and \q{mayaswell2} for all $n,\alpha$ and $\bR$, which implies dim $X_{n,\bR}$ is independent of $n$ and $\bR$.



We would now prove the equivalence of definitions \ref{BR1} and \ref{BR2}. It is sufficient to proof the equivalence for BRs with a single Wyckoff position, since BRs characterized by multiple Wyckoff positions are direct sums of single-Wyckoff BRs.

\noindent \textit{Proof.} 

That definition \ref{BR1} implies definition \ref{BR2} is straightforward: recalling the definition of the site stabilizer $G_{\varpi_n+R}$ as the subgroup of $G$ which preserves $\bvarpi_n{+}\bR$, we may restrict \q{app:Brav2} to $g{=}(p|\bt) {\in} G_{\varpi_n+R}$ by fixing $n'{=}n$ and $\check p \bR{+}\Delta_{g,n',n}{=}\bR$. Then \q{BRW2} follows for $X_{n,\bR}(p) {=} V(p_n^{\mo}pp_n)$. 

Let us now prove the converse: beginning from definition \ref{BR2} and \q{BRW2}, we would  derive \q{app:Brav2} which gives definition \ref{BR1}. 
Utilizing \q{mayaswell2}, we first show that $X_{n,\bR}(p)$ is independent of $\bR$: applying that $G_{\bvarpi_n}$ is conjugate to $G_{\varpi_n+R}{=}(E|\bR) G_{\varpi_n} (E|\bR)^{\mo}$ (cf.\ \q{groupR}), for any  $(p|\boldsymbol{T}){\in}G_{\varpi_n+R}$ there exists $(p|\bS) {\in}G_{\varpi_n}$ such that
\m{ \big[ X_{n,\bR}(p) \big]_{\beta,\alpha} = \braket{ w_{n,\bR}^{\beta}| \widehat{(p|\boldsymbol{T})}| w_{n,\bR}^{\alpha} } \\ 
= \braket{ w_{n,\textbf{0}}^{\beta}| \widehat{(p|\bS)}| w_{n,\textbf{0}}^{\alpha} } =  \big[ X_{n,\textbf{0}}(p) \big]_{\beta,\alpha}. \la{toinsert1}}
 Utilizing \q{mayaswell2},  
one may similarly show that 
\e{X_{n,\textbf{0}}(p){=}X_{1,\textbf{0}}(p_n^{\mo} p p_n).\la{toinsert2}} 
Inserting \qq{toinsert1}{toinsert2} into \q{BRW2},  we derive that for all $g {\in} G_{\varpi_n+R}$,
\e{ \hat g \ket{ w_{n,\bR}^{\alpha} } = \sum_{\beta=1}^{\dim X_{1,\bze}} \big[ X_{1,\bze}(p_n^{\mo}pp_n) \big]_{\beta,\alpha} \ket{ w_{n,\bR}^{\beta} }. \la{BRW23} }
To complete the proof, we would employ a previous observation that $g_{n'}^{\mo} (E|{-}\Delta_{g,n',n}) g g_n\in G_{\varpi}$; this implies that any $g\in G$ can be decomposed into the product $g= (E|\Delta_{g,n',n})g_{n'}hg_n^{\mo}$ for some $h\in G_{\varpi}$. Given the representations of $h$ [cf. \q{BRW23}], $g_n,g_{n'}$ [cf.\ \q{mayaswell2}] and $(E|\Delta_{g,n',n})$ [cf.\ \q{mayaswell1}], we finally derive that  \q{BRW2} implies \q{app:Brav2} with the identification $X_{1,\bze}=V$. This finishes the proof of the equivalence.




\section{Cohomological interpretation of $n(q,p)$}\la{app:remarkcohomological}




In the canonical gauge where $\alpha_p$ is $\bk$-independent, the left-hand-side of \q{alphas} (which vanishes according to \s{proof2}) reduces to
\e{ 2\pi n(q,p) = s_q\alpha_p(\textbf{0}) + \alpha_q(\textbf{0}) - \alpha_{qp}(\textbf{0}). \la{integdef} }
The aim of this appendix is to identify $n(q,p)$ is an inhomogeneous $2$-cocycle, and the equivalence classes of $n(q,p)$ (specified below) as classifying the 1D linear representations   of the point group $\calp$.

To begin, let us review a few notions from group cohomology, as particularized to the present context. For a group $H$, we define the $H$-module as an abelian group on which $H$ acts compatibly with the multiplication operation. In our context, we consider the $\calp$-module $\Z_T$, which equals $\Z$ as a set, and is endowed with an action (denoted $\cdot$) of an element $p{\in}\calp$ on an element $m {\in}\Z_T$: $p {\cdot} m {=} s_p m$.\cite{Chen2013a,Wang2017} The associativity condition $\alpha_{q (p r)}(\textbf{0}){=}\alpha_{(q p) r}(\textbf{0})$ implies that $n(q,p)$ is an inhomogeneous $2$-cocycle, i.e., a map from    from $\calp{\times}\calp$ to $\Z_T$ satisfying that 
\es{ 0 = n(q p,r) + n(q,p) - n(q,p r) - s_q n(p,r); }
the right-hand side may be identified as an inhomogeneous $3$-coboundary. Since $\alpha_p$ is a phase it has a $2\pi$ ambiguity, and $n(q,p)$ is only well-defined modulo: 
\e{  \alpha_p(\textbf{0}) \to \alpha_p(\textbf{0})  + 2\pi \calq(p), \; \calq(p)\in \Z \lin
n(q,p)\rightarrow n(q,p) + s_q \calq(p) + \calq(q) - \calq(qp). \la{cobound} }
The rightmost three terms may be viewed as an inhomogeneous $2$-coboundary. Defining an equivalence for $n(q,p)$ modulo $2$-coboundaries, the equivalence classes of such $n(q,p)$ define\cite{Chen2013a,Rabson2001} the second group cohomology: $H^2(\calp,\Z_T)$; this is isomorphic to $H^1(\calp,U(1)_T)$,\cite{Chen2013a} where $U(1)_T$ is defined as  $U(1)$ that is complex-conjugated under the action of time-reversing elements of $\calp$. In fact $H^1(\calp,U(1)_T)$ is known to classify all the inequivalent 1D linear representations [$e^{i\alpha_p(\boldsymbol{0})} {\in} U(1)$]  of $\calp$.\cite{Chen2013a}




\section{Reality  due to spacetime-inversion symmetry}\la{spacetime}

In a spacetime-inversion symmetric insulator without spin-orbit coupling (i.e., with spin $SU(2)$ symmetry), there exists an antiunitary operator $\tilde T$ which is a symmetry of the Schr\"odinger Hamiltonian $H_0$ and squares to identity. We may work in a spinor basis $f_{s}(\br)$ where $s{=}{\pm} 1$ corresponds to spin up and down in the $z$-direction; in this basis, $\tilde{T}{=}e^{i\pi \sigma_y/2} T\cali $ is the composition of a $\pi$ spin rotation about the $y$-axis (a symmetry of the Hamiltonian), the time-reversal operator  $T {=} \mathrm{e}^{-i \pi \sigma_y/2} K$ (with $K$ implementing complex conjugation and  $T^2 {=} {-}1$) and the spatial-inversion operator $\cali$ which maps $\br {\rightarrow} {-}\br$. In composition, $\tilde{T} {=} K\cali$ preserves the spin component $S_z$, i.e., $\tilde T f_{s}(\br) {=} \bar f_{s}({-}\br)$.

We now study how $\tilde T$ is represented on the Fourier transform of an orthonormal tight-binding basis corresponding to L\"owdin-orthogonalized orbitals $\ket{\phi_{\alpha,\bR}}$, where $\alpha {=} 1,{\ldots},N$ is orbital index and $\bR$ a BL, i.e., on
\e{ \ket{ u_{\alpha,\bk} }_{\text{cell}} = \f{1}{\sqrt{|T^d|}} \sum_{\bR} \mathrm{e}^{i \bk \cdot (\bR-\br)} \ket{ \phi_{\alpha,\bR}}_{\text{cell}}. \la{blochdef} }
$\braket{\,.\,|\,.\,}_{\text{cell}}$ denotes the inner product over one unit cell. The tight-binding Hamiltonian is defined as the matrix elements of the Hamiltonian $H_0$ in this basis:
\e{ [ h(\bk) ]_{\ab} = \braopket{u_{\alpha,\bk}}{e^{-i\bk\cdot \br}H_0\eikr}{u_{\beta,\bk}}_{\text{cell}}.}
The unitary component of the representation of $\tilde T$ is\cite{Alexandradinata2016}
\e{ [\mathcal{B}(\bk)]_{\beta \alpha} = \braket{ u_{\beta, \bk} | \tilde T | u_{\alpha,\bk} }_{\text{cell}} = \int_{\text{cell}} \bar u_{\beta,\bk}(\br) \bar u_{\alpha,\bk}(-\br) \mathrm d\br; }
one may verify that this matrix is unitary:
\e{ \sum_{\gamma=1}^N[\mathcal{B}(\bk)]_{\alpha \gamma} [\bar{\mathcal{B}}(\bk)]_{\beta \gamma } = \braket{ u_{\alpha, \bk} | u_{\beta,\bk} }_{\text{cell}} = \delta_{\alpha, \beta}, }
and symmetric: $[\mathcal{B}(\bk)]_{\beta \alpha} {=} [\mathcal{B}(\bk)]_{\alpha \beta}$.


We now show that there exists a basis of $\ket{ u_{\alpha,\bk} }$ for which $h(\bk)$ is real and symmetric at each $\bk$.  $[\tilde T,H]{=}0$ implies that
\e{ h(\bk)_{\beta \alpha} &= \braket{ u_{\beta,\bk}| \mathrm{e}^{-i \bk \cdot \br} H_0 \mathrm{e}^{i \bk \cdot \br} | u_{\alpha,\bk} }_{\text{cell}} \lin 
&= \braket{ u_{\beta,\bk}| \tilde T \mathrm{e}^{-i \bk \cdot \br} H_0 \mathrm{e}^{i \bk \cdot \br} \tilde T | u_{\alpha,\bk} }_{\text{cell}} \lin
&= [\mathcal{B}(\bk)]_{\beta \gamma} \bar{h}(\bk)_{\gamma \delta} [\bar{\mathcal{B}}(\bk)]_{\delta \alpha} \lin 
&= [ \mathcal{B}(\bk) \bar{h}(\bk) \bar{\mathcal{B}}(\bk) ]_{\beta,\alpha}, \la{cham}}
using the Einstein summation convention.
The symmetric matrix $\mathcal{B}(\bk)$ can be diagonalized by an orthogonal matrix $S(\bk)$: $\mathcal{B}(\bk) {=} S(\bk) D(\bk) S^T(\bk)$ where $^T$ denotes transposition. $D$ is a diagonal matrix with unimodular diagonal entries, and therefore $D^{\mo}(\bk) {=} \bar{D}(\bk)$. We now implement the unitary transformation $U(\bk) {=} S(\bk) D^{1/2}(\bk)$, \cite{Fang2015S}  such that the transformed Hamiltonian is real at each $\bk$:
\e{ h' :=&\; U^{\dagger} h U=  D^{-1/2} S^T \big( S D S^T \bar{h} S \bar D S^T \big) S D^{1/2} \lin
=&\; D^{1/2} S^T \bar{h} S D^{-1/2} = \bar{h}'. }

\bibliography{EBR}

\begin{thebibliography}{118}%
\makeatletter
\providecommand \@ifxundefined [1]{%
 \@ifx{#1\undefined}
}%
\providecommand \@ifnum [1]{%
 \ifnum #1\expandafter \@firstoftwo
 \else \expandafter \@secondoftwo
 \fi
}%
\providecommand \@ifx [1]{%
 \ifx #1\expandafter \@firstoftwo
 \else \expandafter \@secondoftwo
 \fi
}%
\providecommand \natexlab [1]{#1}%
\providecommand \enquote  [1]{``#1''}%
\providecommand \bibnamefont  [1]{#1}%
\providecommand \bibfnamefont [1]{#1}%
\providecommand \citenamefont [1]{#1}%
\providecommand \href@noop [0]{\@secondoftwo}%
\providecommand \href [0]{\begingroup \@sanitize@url \@href}%
\providecommand \@href[1]{\@@startlink{#1}\@@href}%
\providecommand \@@href[1]{\endgroup#1\@@endlink}%
\providecommand \@sanitize@url [0]{\catcode `\\12\catcode `\$12\catcode
  `\&12\catcode `\#12\catcode `\^12\catcode `\_12\catcode `\%12\relax}%
\providecommand \@@startlink[1]{}%
\providecommand \@@endlink[0]{}%
\providecommand \url  [0]{\begingroup\@sanitize@url \@url }%
\providecommand \@url [1]{\endgroup\@href {#1}{\urlprefix }}%
\providecommand \urlprefix  [0]{URL }%
\providecommand \Eprint [0]{\href }%
\providecommand \doibase [0]{http://dx.doi.org/}%
\providecommand \selectlanguage [0]{\@gobble}%
\providecommand \bibinfo  [0]{\@secondoftwo}%
\providecommand \bibfield  [0]{\@secondoftwo}%
\providecommand \translation [1]{[#1]}%
\providecommand \BibitemOpen [0]{}%
\providecommand \bibitemStop [0]{}%
\providecommand \bibitemNoStop [0]{.\EOS\space}%
\providecommand \EOS [0]{\spacefactor3000\relax}%
\providecommand \BibitemShut  [1]{\csname bibitem#1\endcsname}%
\let\auto@bib@innerbib\@empty
\bibitem [{\citenamefont {Resta}(1994)}]{Resta1994}%
  \BibitemOpen
  \bibfield  {author} {\bibinfo {author} {\bibfnamefont {R.}~\bibnamefont
  {Resta}},\ }\href {\doibase 10.1103/RevModPhys.66.899} {\bibfield  {journal}
  {\bibinfo  {journal} {Rev. Mod. Phys.}\ }\textbf {\bibinfo {volume} {66}},\
  \bibinfo {pages} {899} (\bibinfo {year} {1994})}\BibitemShut {NoStop}%
\bibitem [{\citenamefont {Soluyanov}\ and\ \citenamefont
  {Vanderbilt}(2011)}]{Soluyanov2011}%
  \BibitemOpen
  \bibfield  {author} {\bibinfo {author} {\bibfnamefont {A.~A.}\ \bibnamefont
  {Soluyanov}}\ and\ \bibinfo {author} {\bibfnamefont {D.}~\bibnamefont
  {Vanderbilt}},\ }\href {\doibase 10.1103/PhysRevB.83.035108} {\bibfield
  {journal} {\bibinfo  {journal} {Phys. Rev. B}\ }\textbf {\bibinfo {volume}
  {83}},\ \bibinfo {pages} {035108} (\bibinfo {year} {2011})}\BibitemShut
  {NoStop}%
\bibitem [{\citenamefont {Taherinejad}\ \emph {et~al.}(2014)\citenamefont
  {Taherinejad}, \citenamefont {Garrity},\ and\ \citenamefont
  {Vanderbilt}}]{Maryam2014}%
  \BibitemOpen
  \bibfield  {author} {\bibinfo {author} {\bibfnamefont {M.}~\bibnamefont
  {Taherinejad}}, \bibinfo {author} {\bibfnamefont {K.~F.}\ \bibnamefont
  {Garrity}}, \ and\ \bibinfo {author} {\bibfnamefont {D.}~\bibnamefont
  {Vanderbilt}},\ }\href {\doibase 10.1103/PhysRevB.89.115102} {\bibfield
  {journal} {\bibinfo  {journal} {Phys. Rev. B}\ }\textbf {\bibinfo {volume}
  {89}},\ \bibinfo {pages} {115102} (\bibinfo {year} {2014})}\BibitemShut
  {NoStop}%
\bibitem [{\citenamefont {Read}(2017)}]{Read2017}%
  \BibitemOpen
  \bibfield  {author} {\bibinfo {author} {\bibfnamefont {N.}~\bibnamefont
  {Read}},\ }\href {\doibase 10.1103/PhysRevB.95.115309} {\bibfield  {journal}
  {\bibinfo  {journal} {Phys. Rev. B}\ }\textbf {\bibinfo {volume} {95}},\
  \bibinfo {pages} {115309} (\bibinfo {year} {2017})}\BibitemShut {NoStop}%
\bibitem [{\citenamefont {Bradlyn}\ \emph {et~al.}(2017)\citenamefont
  {Bradlyn}, \citenamefont {Elcoro}, \citenamefont {Cano}, \citenamefont
  {Vergniory}, \citenamefont {Wang}, \citenamefont {Felser}, \citenamefont
  {Aroyo},\ and\ \citenamefont {Bernevig}}]{Bradlyn2017}%
  \BibitemOpen
  \bibfield  {author} {\bibinfo {author} {\bibfnamefont {B.}~\bibnamefont
  {Bradlyn}}, \bibinfo {author} {\bibfnamefont {L.}~\bibnamefont {Elcoro}},
  \bibinfo {author} {\bibfnamefont {J.}~\bibnamefont {Cano}}, \bibinfo {author}
  {\bibfnamefont {M.~G.}\ \bibnamefont {Vergniory}}, \bibinfo {author}
  {\bibfnamefont {Z.}~\bibnamefont {Wang}}, \bibinfo {author} {\bibfnamefont
  {C.}~\bibnamefont {Felser}}, \bibinfo {author} {\bibfnamefont {M.~I.}\
  \bibnamefont {Aroyo}}, \ and\ \bibinfo {author} {\bibfnamefont {B.~A.}\
  \bibnamefont {Bernevig}},\ }\href@noop {} {\bibfield  {journal} {\bibinfo
  {journal} {Nature}\ }\textbf {\bibinfo {volume} {547}},\ \bibinfo {pages}
  {298} (\bibinfo {year} {2017})}\BibitemShut {NoStop}%
\bibitem [{\citenamefont {Cano}\ \emph {et~al.}(2018)\citenamefont {Cano},
  \citenamefont {Bradlyn}, \citenamefont {Wang}, \citenamefont {Elcoro},
  \citenamefont {Vergniory}, \citenamefont {Felser}, \citenamefont {Aroyo},\
  and\ \citenamefont {Bernevig}}]{Cano2017}%
  \BibitemOpen
  \bibfield  {author} {\bibinfo {author} {\bibfnamefont {J.}~\bibnamefont
  {Cano}}, \bibinfo {author} {\bibfnamefont {B.}~\bibnamefont {Bradlyn}},
  \bibinfo {author} {\bibfnamefont {Z.}~\bibnamefont {Wang}}, \bibinfo {author}
  {\bibfnamefont {L.}~\bibnamefont {Elcoro}}, \bibinfo {author} {\bibfnamefont
  {M.~G.}\ \bibnamefont {Vergniory}}, \bibinfo {author} {\bibfnamefont
  {C.}~\bibnamefont {Felser}}, \bibinfo {author} {\bibfnamefont {M.~I.}\
  \bibnamefont {Aroyo}}, \ and\ \bibinfo {author} {\bibfnamefont {B.~A.}\
  \bibnamefont {Bernevig}},\ }\href {\doibase 10.1103/PhysRevB.97.035139}
  {\bibfield  {journal} {\bibinfo  {journal} {Phys. Rev. B}\ }\textbf {\bibinfo
  {volume} {97}},\ \bibinfo {pages} {035139} (\bibinfo {year}
  {2018})}\BibitemShut {NoStop}%
\bibitem [{\citenamefont {Hoeller}\ and\ \citenamefont
  {Alexandradinata}(2017)}]{Holler2017}%
  \BibitemOpen
  \bibfield  {author} {\bibinfo {author} {\bibfnamefont {J.}~\bibnamefont
  {Hoeller}}\ and\ \bibinfo {author} {\bibfnamefont {A.}~\bibnamefont
  {Alexandradinata}},\ }\href {https://arxiv.org/abs/1708.02943} {\bibfield
  {journal} {\bibinfo  {journal} {arXiv}\ } (\bibinfo {year} {2017})},\ \Eprint
  {http://arxiv.org/abs/1708.02943} {arXiv:1708.02943} \BibitemShut {NoStop}%
\bibitem [{\citenamefont {Thouless}(1984)}]{Thouless1984}%
  \BibitemOpen
  \bibfield  {author} {\bibinfo {author} {\bibfnamefont {D.~J.}\ \bibnamefont
  {Thouless}},\ }\href {\doibase 10.1088/0022-3719/17/12/003} {\bibfield
  {journal} {\bibinfo  {journal} {J. Phys. C Solid State Phys.}\ }\textbf
  {\bibinfo {volume} {17}},\ \bibinfo {pages} {L325} (\bibinfo {year}
  {1984})}\BibitemShut {NoStop}%
\bibitem [{\citenamefont {Thonhauser}\ and\ \citenamefont
  {Vanderbilt}(2006)}]{Thonhauser2006}%
  \BibitemOpen
  \bibfield  {author} {\bibinfo {author} {\bibfnamefont {T.}~\bibnamefont
  {Thonhauser}}\ and\ \bibinfo {author} {\bibfnamefont {D.}~\bibnamefont
  {Vanderbilt}},\ }\href {\doibase 10.1103/PhysRevB.74.235111} {\bibfield
  {journal} {\bibinfo  {journal} {Phys. Rev. B}\ }\textbf {\bibinfo {volume}
  {74}},\ \bibinfo {pages} {235111} (\bibinfo {year} {2006})}\BibitemShut
  {NoStop}%
\bibitem [{\citenamefont {Brouder}\ \emph {et~al.}(2007)\citenamefont
  {Brouder}, \citenamefont {Panati}, \citenamefont {Calandra}, \citenamefont
  {Mourougane},\ and\ \citenamefont {Marzari}}]{Brouder2007}%
  \BibitemOpen
  \bibfield  {author} {\bibinfo {author} {\bibfnamefont {C.}~\bibnamefont
  {Brouder}}, \bibinfo {author} {\bibfnamefont {G.}~\bibnamefont {Panati}},
  \bibinfo {author} {\bibfnamefont {M.}~\bibnamefont {Calandra}}, \bibinfo
  {author} {\bibfnamefont {C.}~\bibnamefont {Mourougane}}, \ and\ \bibinfo
  {author} {\bibfnamefont {N.}~\bibnamefont {Marzari}},\ }\href {\doibase
  10.1103/PhysRevLett.98.046402} {\bibfield  {journal} {\bibinfo  {journal}
  {Phys. Rev. Lett.}\ }\textbf {\bibinfo {volume} {98}},\ \bibinfo {pages}
  {046402} (\bibinfo {year} {2007})}\BibitemShut {NoStop}%
\bibitem [{\citenamefont {Budich}\ \emph {et~al.}(2014)\citenamefont {Budich},
  \citenamefont {Eisert}, \citenamefont {Bergholtz}, \citenamefont {Diehl},\
  and\ \citenamefont {Zoller}}]{Budich2014}%
  \BibitemOpen
  \bibfield  {author} {\bibinfo {author} {\bibfnamefont {J.~C.}\ \bibnamefont
  {Budich}}, \bibinfo {author} {\bibfnamefont {J.}~\bibnamefont {Eisert}},
  \bibinfo {author} {\bibfnamefont {E.~J.}\ \bibnamefont {Bergholtz}}, \bibinfo
  {author} {\bibfnamefont {S.}~\bibnamefont {Diehl}}, \ and\ \bibinfo {author}
  {\bibfnamefont {P.}~\bibnamefont {Zoller}},\ }\href {\doibase
  10.1103/PhysRevB.90.115110} {\bibfield  {journal} {\bibinfo  {journal} {Phys.
  Rev. B}\ }\textbf {\bibinfo {volume} {90}},\ \bibinfo {pages} {115110}
  (\bibinfo {year} {2014})}\BibitemShut {NoStop}%
\bibitem [{\citenamefont {Kohn}(1973)}]{Kohn1973}%
  \BibitemOpen
  \bibfield  {author} {\bibinfo {author} {\bibfnamefont {W.}~\bibnamefont
  {Kohn}},\ }\href {\doibase 10.1103/PhysRevB.7.4388} {\bibfield  {journal}
  {\bibinfo  {journal} {Phys. Rev. B}\ }\textbf {\bibinfo {volume} {7}},\
  \bibinfo {pages} {4388} (\bibinfo {year} {1973})}\BibitemShut {NoStop}%
\bibitem [{\citenamefont {Cloizeaux}(1964)}]{Cloizeaux1964}%
  \BibitemOpen
  \bibfield  {author} {\bibinfo {author} {\bibfnamefont {J.~D.}\ \bibnamefont
  {Cloizeaux}},\ }\href@noop {} {\bibfield  {journal} {\bibinfo  {journal}
  {Phys. Rev.}\ }\textbf {\bibinfo {volume} {135}},\ \bibinfo {pages} {685}
  (\bibinfo {year} {1964})}\BibitemShut {NoStop}%
\bibitem [{\citenamefont {Nenciu}(1983)}]{Nenciu1983}%
  \BibitemOpen
  \bibfield  {author} {\bibinfo {author} {\bibfnamefont {G.}~\bibnamefont
  {Nenciu}},\ }\href {https://projecteuclid.org/euclid.cmp/1103940475}
  {\bibfield  {journal} {\bibinfo  {journal} {Commun. Math. Phys.}\ }\textbf
  {\bibinfo {volume} {91}},\ \bibinfo {pages} {81} (\bibinfo {year}
  {1983})}\BibitemShut {NoStop}%
\bibitem [{\citenamefont {\textrm{C. L. Kane}}\ and\ \citenamefont {\textrm{E.
  J. Mele}}(2005)}]{kane2005A}%
  \BibitemOpen
  \bibfield  {author} {\bibinfo {author} {\bibnamefont {\textrm{C. L. Kane}}}\
  and\ \bibinfo {author} {\bibnamefont {\textrm{E. J. Mele}}},\ }\href@noop {}
  {\bibfield  {journal} {\bibinfo  {journal} {Phys. Rev. Lett.}\ }\textbf
  {\bibinfo {volume} {95}},\ \bibinfo {pages} {226801} (\bibinfo {year}
  {2005})}\BibitemShut {NoStop}%
\bibitem [{\citenamefont {Kane}\ and\ \citenamefont {Mele}(2005)}]{Kane2005b}%
  \BibitemOpen
  \bibfield  {author} {\bibinfo {author} {\bibfnamefont {C.~L.}\ \bibnamefont
  {Kane}}\ and\ \bibinfo {author} {\bibfnamefont {E.~J.}\ \bibnamefont
  {Mele}},\ }\href {\doibase 10.1103/PhysRevLett.95.146802} {\bibfield
  {journal} {\bibinfo  {journal} {Phys. Rev. Lett.}\ }\textbf {\bibinfo
  {volume} {95}},\ \bibinfo {pages} {146802} (\bibinfo {year}
  {2005})}\BibitemShut {NoStop}%
\bibitem [{\citenamefont {\textrm{B.A. Bernevig}}\ and\ \citenamefont
  {\textrm{S.C. Zhang}}(2006)}]{QSHE_bernevig}%
  \BibitemOpen
  \bibfield  {author} {\bibinfo {author} {\bibnamefont {\textrm{B.A.
  Bernevig}}}\ and\ \bibinfo {author} {\bibnamefont {\textrm{S.C. Zhang}}},\
  }\href@noop {} {\bibfield  {journal} {\bibinfo  {journal} {Phys. Rev. Lett.}\
  }\textbf {\bibinfo {volume} {96}},\ \bibinfo {pages} {106802} (\bibinfo
  {year} {2006})}\BibitemShut {NoStop}%
\bibitem [{\citenamefont {\textrm{B. A. Bernevig}}\ \emph
  {et~al.}(2006)\citenamefont {\textrm{B. A. Bernevig}}, \citenamefont
  {\textrm{T. L. Hughes}},\ and\ \citenamefont {\textrm{S.C.
  Zhang}}}]{HgTe_bernevig}%
  \BibitemOpen
  \bibfield  {author} {\bibinfo {author} {\bibnamefont {\textrm{B. A.
  Bernevig}}}, \bibinfo {author} {\bibnamefont {\textrm{T. L. Hughes}}}, \ and\
  \bibinfo {author} {\bibnamefont {\textrm{S.C. Zhang}}},\ }\href@noop {}
  {\bibfield  {journal} {\bibinfo  {journal} {Science}\ }\textbf {\bibinfo
  {volume} {314}},\ \bibinfo {pages} {1757} (\bibinfo {year}
  {2006})}\BibitemShut {NoStop}%
\bibitem [{\citenamefont {Roy}(2009{\natexlab{a}})}]{QSHE_Rahul}%
  \BibitemOpen
  \bibfield  {author} {\bibinfo {author} {\bibfnamefont {R.}~\bibnamefont
  {Roy}},\ }\href@noop {} {\bibfield  {journal} {\bibinfo  {journal} {Phys.
  Rev. B}\ }\textbf {\bibinfo {volume} {79}},\ \bibinfo {pages} {195321}
  (\bibinfo {year} {2009}{\natexlab{a}})}\BibitemShut {NoStop}%
\bibitem [{\citenamefont {Fu}\ and\ \citenamefont {Kane}(2006)}]{fu2006}%
  \BibitemOpen
  \bibfield  {author} {\bibinfo {author} {\bibfnamefont {L.}~\bibnamefont
  {Fu}}\ and\ \bibinfo {author} {\bibfnamefont {C.~L.}\ \bibnamefont {Kane}},\
  }\href {\doibase 10.1103/PhysRevB.74.195312} {\bibfield  {journal} {\bibinfo
  {journal} {Phys. Rev. B}\ }\textbf {\bibinfo {volume} {74}},\ \bibinfo {eid}
  {195312} (\bibinfo {year} {2006})}\BibitemShut {NoStop}%
\bibitem [{\citenamefont {Dyson}(1962)}]{Dyson1962}%
  \BibitemOpen
  \bibfield  {author} {\bibinfo {author} {\bibfnamefont {F.~J.}\ \bibnamefont
  {Dyson}},\ }\href {\doibase 10.1063/1.1703863} {\bibfield  {journal}
  {\bibinfo  {journal} {J. Math. Phys.}\ }\textbf {\bibinfo {volume} {3}},\
  \bibinfo {pages} {1199} (\bibinfo {year} {1962})}\BibitemShut {NoStop}%
\bibitem [{Note1()}]{Note1}%
  \BibitemOpen
  \bibinfo {note} {Our results do not immediately apply to band topological
  superconductors, which have an additional particle-hole and/or chiral
  symmetries in the Bogoliubov-de-Gennes formalism.\cite
  {schnyder_classify3DTIandTSC} The possibility to formulate band
  representations for superconductors remains an open and interesting
  question.}\BibitemShut {Stop}%
\bibitem [{\citenamefont {Bradley}\ and\ \citenamefont
  {Davies}(1968)}]{magnetic_groups}%
  \BibitemOpen
  \bibfield  {author} {\bibinfo {author} {\bibfnamefont {C.~J.}\ \bibnamefont
  {Bradley}}\ and\ \bibinfo {author} {\bibfnamefont {B.~L.}\ \bibnamefont
  {Davies}},\ }\href@noop {} {\bibfield  {journal} {\bibinfo  {journal} {Rev.
  Mod. Phys.}\ }\textbf {\bibinfo {volume} {40}},\ \bibinfo {pages} {359}
  (\bibinfo {year} {1968})}\BibitemShut {NoStop}%
\bibitem [{\citenamefont {Watanabe}\ \emph {et~al.}(2017)\citenamefont
  {Watanabe}, \citenamefont {Po},\ and\ \citenamefont
  {Vishwanath}}]{Watanabe2017a}%
  \BibitemOpen
  \bibfield  {author} {\bibinfo {author} {\bibfnamefont {H.}~\bibnamefont
  {Watanabe}}, \bibinfo {author} {\bibfnamefont {H.~C.}\ \bibnamefont {Po}}, \
  and\ \bibinfo {author} {\bibfnamefont {A.}~\bibnamefont {Vishwanath}},\
  }\href {http://arxiv.org/abs/1707.01903} {\bibfield  {journal} {\bibinfo
  {journal} {arXiv}\ } (\bibinfo {year} {2017})},\ \Eprint
  {http://arxiv.org/abs/1707.01903} {arXiv:1707.01903} \BibitemShut {NoStop}%
\bibitem [{\citenamefont {Panati}(2007)}]{Panati2007}%
  \BibitemOpen
  \bibfield  {author} {\bibinfo {author} {\bibfnamefont {G.}~\bibnamefont
  {Panati}},\ }\href {\doibase 10.1007/s00023-007-0326-8} {\bibfield  {journal}
  {\bibinfo  {journal} {Ann. Henri Poincar{\'{e}}}\ }\textbf {\bibinfo {volume}
  {8}},\ \bibinfo {pages} {995} (\bibinfo {year} {2007})}\BibitemShut {NoStop}%
\bibitem [{\citenamefont {Zhang}\ \emph {et~al.}(2009)\citenamefont {Zhang},
  \citenamefont {Liu}, \citenamefont {Qi}, \citenamefont {Dai}, \citenamefont
  {Fang},\ and\ \citenamefont {Zhang}}]{zhang2009}%
  \BibitemOpen
  \bibfield  {author} {\bibinfo {author} {\bibfnamefont {H.}~\bibnamefont
  {Zhang}}, \bibinfo {author} {\bibfnamefont {C.-X.}\ \bibnamefont {Liu}},
  \bibinfo {author} {\bibfnamefont {X.-L.}\ \bibnamefont {Qi}}, \bibinfo
  {author} {\bibfnamefont {X.}~\bibnamefont {Dai}}, \bibinfo {author}
  {\bibfnamefont {Z.}~\bibnamefont {Fang}}, \ and\ \bibinfo {author}
  {\bibfnamefont {S.-C.}\ \bibnamefont {Zhang}},\ }\href@noop {} {\bibfield
  {journal} {\bibinfo  {journal} {Nat. Phys.}\ }\textbf {\bibinfo {volume}
  {5}},\ \bibinfo {pages} {438} (\bibinfo {year} {2009})}\BibitemShut {NoStop}%
\bibitem [{\citenamefont {Hsieh}\ \emph {et~al.}(2008)\citenamefont {Hsieh},
  \citenamefont {Qian}, \citenamefont {Wray}, \citenamefont {Xia},
  \citenamefont {Hor}, \citenamefont {Cava},\ and\ \citenamefont
  {Hasan}}]{hsieh2008}%
  \BibitemOpen
  \bibfield  {author} {\bibinfo {author} {\bibfnamefont {D.}~\bibnamefont
  {Hsieh}}, \bibinfo {author} {\bibfnamefont {D.}~\bibnamefont {Qian}},
  \bibinfo {author} {\bibfnamefont {L.}~\bibnamefont {Wray}}, \bibinfo {author}
  {\bibfnamefont {Y.}~\bibnamefont {Xia}}, \bibinfo {author} {\bibfnamefont
  {Y.~S.}\ \bibnamefont {Hor}}, \bibinfo {author} {\bibfnamefont {R.~J.}\
  \bibnamefont {Cava}}, \ and\ \bibinfo {author} {\bibfnamefont {M.~Z.}\
  \bibnamefont {Hasan}},\ }\href@noop {} {\bibfield  {journal} {\bibinfo
  {journal} {Nature}\ }\textbf {\bibinfo {volume} {452}},\ \bibinfo {pages}
  {970} (\bibinfo {year} {2008})}\BibitemShut {NoStop}%
\bibitem [{\citenamefont {Fu}\ \emph {et~al.}(2007)\citenamefont {Fu},
  \citenamefont {Kane},\ and\ \citenamefont {Mele}}]{fukanemele_3DTI}%
  \BibitemOpen
  \bibfield  {author} {\bibinfo {author} {\bibfnamefont {L.}~\bibnamefont
  {Fu}}, \bibinfo {author} {\bibfnamefont {C.~L.}\ \bibnamefont {Kane}}, \ and\
  \bibinfo {author} {\bibfnamefont {E.~J.}\ \bibnamefont {Mele}},\ }\href
  {\doibase 10.1103/PhysRevLett.98.106803} {\bibfield  {journal} {\bibinfo
  {journal} {Phys. Rev. Lett.}\ }\textbf {\bibinfo {volume} {98}},\ \bibinfo
  {eid} {106803} (\bibinfo {year} {2007})}\BibitemShut {NoStop}%
\bibitem [{\citenamefont {Fu}\ and\ \citenamefont {Kane}(2007)}]{Inversion_Fu}%
  \BibitemOpen
  \bibfield  {author} {\bibinfo {author} {\bibfnamefont {L.}~\bibnamefont
  {Fu}}\ and\ \bibinfo {author} {\bibfnamefont {C.~L.}\ \bibnamefont {Kane}},\
  }\href@noop {} {\bibfield  {journal} {\bibinfo  {journal} {Phys. Rev. B}\
  }\textbf {\bibinfo {volume} {76}},\ \bibinfo {pages} {045302} (\bibinfo
  {year} {2007})}\BibitemShut {NoStop}%
\bibitem [{\citenamefont {Moore}\ and\ \citenamefont
  {Balents}(2007)}]{moore2007}%
  \BibitemOpen
  \bibfield  {author} {\bibinfo {author} {\bibfnamefont {J.~E.}\ \bibnamefont
  {Moore}}\ and\ \bibinfo {author} {\bibfnamefont {L.}~\bibnamefont
  {Balents}},\ }\href {\doibase 10.1103/PhysRevB.75.121306} {\bibfield
  {journal} {\bibinfo  {journal} {Phys. Rev. B}\ }\textbf {\bibinfo {volume}
  {75}},\ \bibinfo {eid} {121306} (\bibinfo {year} {2007})}\BibitemShut
  {NoStop}%
\bibitem [{\citenamefont {Roy}(2009{\natexlab{b}})}]{Rahul_3DTI}%
  \BibitemOpen
  \bibfield  {author} {\bibinfo {author} {\bibfnamefont {R.}~\bibnamefont
  {Roy}},\ }\href {\doibase 10.1103/PhysRevB.79.195322} {\bibfield  {journal}
  {\bibinfo  {journal} {Phys. Rev. B}\ }\textbf {\bibinfo {volume} {79}},\
  \bibinfo {pages} {195322} (\bibinfo {year} {2009}{\natexlab{b}})}\BibitemShut
  {NoStop}%
\bibitem [{\citenamefont {Hsieh}\ \emph {et~al.}(2012)\citenamefont {Hsieh},
  \citenamefont {Lin}, \citenamefont {Liu}, \citenamefont {Duan}, \citenamefont
  {Bansil},\ and\ \citenamefont {Fu}}]{Hsieh_SnTe}%
  \BibitemOpen
  \bibfield  {author} {\bibinfo {author} {\bibfnamefont {T.~H.}\ \bibnamefont
  {Hsieh}}, \bibinfo {author} {\bibfnamefont {H.}~\bibnamefont {Lin}}, \bibinfo
  {author} {\bibfnamefont {J.}~\bibnamefont {Liu}}, \bibinfo {author}
  {\bibfnamefont {W.}~\bibnamefont {Duan}}, \bibinfo {author} {\bibfnamefont
  {A.}~\bibnamefont {Bansil}}, \ and\ \bibinfo {author} {\bibfnamefont
  {L.}~\bibnamefont {Fu}},\ }\href@noop {} {\bibfield  {journal} {\bibinfo
  {journal} {Nat. Comm.}\ }\textbf {\bibinfo {volume} {3}},\ \bibinfo {pages}
  {982} (\bibinfo {year} {2012})}\BibitemShut {NoStop}%
\bibitem [{\citenamefont {Xu}(2012)}]{Xu_observeSnTe}%
  \BibitemOpen
  \bibfield  {author} {\bibinfo {author} {\bibfnamefont {S.-Y.}\ \bibnamefont
  {Xu}},\ }\href@noop {} {\bibfield  {journal} {\bibinfo  {journal} {et al.,
  Nat. Commun. 3:1192 doi: 10.1038/ncomms2191}\ } (\bibinfo {year}
  {2012})}\BibitemShut {NoStop}%
\bibitem [{\citenamefont {Tanaka}(2012)}]{Tanaka_observeSnTe}%
  \BibitemOpen
  \bibfield  {author} {\bibinfo {author} {\bibfnamefont {Y.}~\bibnamefont
  {Tanaka}},\ }\href@noop {} {\bibfield  {journal} {\bibinfo  {journal} {et
  al., Nature Physics}\ }\textbf {\bibinfo {volume} {8}},\ \bibinfo {pages}
  {800} (\bibinfo {year} {2012})}\BibitemShut {NoStop}%
\bibitem [{\citenamefont {Alexandradinata}\ \emph
  {et~al.}(2016{\natexlab{a}})\citenamefont {Alexandradinata}, \citenamefont
  {Wang},\ and\ \citenamefont {Bernevig}}]{Hourglass2016}%
  \BibitemOpen
  \bibfield  {author} {\bibinfo {author} {\bibfnamefont {A.}~\bibnamefont
  {Alexandradinata}}, \bibinfo {author} {\bibfnamefont {Z.}~\bibnamefont
  {Wang}}, \ and\ \bibinfo {author} {\bibfnamefont {B.~A.}\ \bibnamefont
  {Bernevig}},\ }\href@noop {} {\bibfield  {journal} {\bibinfo  {journal}
  {Nature}\ }\textbf {\bibinfo {volume} {532}},\ \bibinfo {pages} {189}
  (\bibinfo {year} {2016}{\natexlab{a}})}\BibitemShut {NoStop}%
\bibitem [{\citenamefont {Alexandradinata}\ \emph
  {et~al.}(2016{\natexlab{b}})\citenamefont {Alexandradinata}, \citenamefont
  {Wang},\ and\ \citenamefont {Bernevig}}]{Alexandradinata2016}%
  \BibitemOpen
  \bibfield  {author} {\bibinfo {author} {\bibfnamefont {A.}~\bibnamefont
  {Alexandradinata}}, \bibinfo {author} {\bibfnamefont {Z.}~\bibnamefont
  {Wang}}, \ and\ \bibinfo {author} {\bibfnamefont {B.~A.}\ \bibnamefont
  {Bernevig}},\ }\href {\doibase 10.1103/PhysRevX.6.021008} {\bibfield
  {journal} {\bibinfo  {journal} {Phys. Rev. X}\ }\textbf {\bibinfo {volume}
  {6}},\ \bibinfo {pages} {021008} (\bibinfo {year}
  {2016}{\natexlab{b}})}\BibitemShut {NoStop}%
\bibitem [{\citenamefont {Ma}\ \emph {et~al.}(2017)\citenamefont {Ma},
  \citenamefont {Yi}, \citenamefont {Lv}, \citenamefont {Wang}, \citenamefont
  {Nie}, \citenamefont {Wang}, \citenamefont {Kong}, \citenamefont {Huang},
  \citenamefont {Richard}, \citenamefont {Zhang}, \citenamefont {Yaji},
  \citenamefont {Kuroda}, \citenamefont {Shin}, \citenamefont {Weng},
  \citenamefont {Bernevig}, \citenamefont {Shi}, \citenamefont {Qian},\ and\
  \citenamefont {Ding}}]{Ma_discoverhourglass}%
  \BibitemOpen
  \bibfield  {author} {\bibinfo {author} {\bibfnamefont {J.}~\bibnamefont
  {Ma}}, \bibinfo {author} {\bibfnamefont {C.}~\bibnamefont {Yi}}, \bibinfo
  {author} {\bibfnamefont {B.}~\bibnamefont {Lv}}, \bibinfo {author}
  {\bibfnamefont {Z.}~\bibnamefont {Wang}}, \bibinfo {author} {\bibfnamefont
  {S.}~\bibnamefont {Nie}}, \bibinfo {author} {\bibfnamefont {L.}~\bibnamefont
  {Wang}}, \bibinfo {author} {\bibfnamefont {L.}~\bibnamefont {Kong}}, \bibinfo
  {author} {\bibfnamefont {Y.}~\bibnamefont {Huang}}, \bibinfo {author}
  {\bibfnamefont {P.}~\bibnamefont {Richard}}, \bibinfo {author} {\bibfnamefont
  {P.}~\bibnamefont {Zhang}}, \bibinfo {author} {\bibfnamefont
  {K.}~\bibnamefont {Yaji}}, \bibinfo {author} {\bibfnamefont {K.}~\bibnamefont
  {Kuroda}}, \bibinfo {author} {\bibfnamefont {S.}~\bibnamefont {Shin}},
  \bibinfo {author} {\bibfnamefont {H.}~\bibnamefont {Weng}}, \bibinfo {author}
  {\bibfnamefont {B.~A.}\ \bibnamefont {Bernevig}}, \bibinfo {author}
  {\bibfnamefont {Y.}~\bibnamefont {Shi}}, \bibinfo {author} {\bibfnamefont
  {T.}~\bibnamefont {Qian}}, \ and\ \bibinfo {author} {\bibfnamefont
  {H.}~\bibnamefont {Ding}},\ }\href {\doibase 10.1126/sciadv.1602415}
  {\bibfield  {journal} {\bibinfo  {journal} {Science Advances}\ }\textbf
  {\bibinfo {volume} {3}} (\bibinfo {year} {2017}),\
  10.1126/sciadv.1602415}\BibitemShut {NoStop}%
\bibitem [{\citenamefont {Fu}(2011)}]{Fu2011}%
  \BibitemOpen
  \bibfield  {author} {\bibinfo {author} {\bibfnamefont {L.}~\bibnamefont
  {Fu}},\ }\href {\doibase 10.1103/PhysRevLett.106.106802} {\bibfield
  {journal} {\bibinfo  {journal} {Phys. Rev. Lett.}\ }\textbf {\bibinfo
  {volume} {106}},\ \bibinfo {pages} {106802} (\bibinfo {year}
  {2011})}\BibitemShut {NoStop}%
\bibitem [{\citenamefont {Liu}\ \emph {et~al.}(2014)\citenamefont {Liu},
  \citenamefont {Zhang},\ and\ \citenamefont {VanLeeuwen}}]{Liu2014}%
  \BibitemOpen
  \bibfield  {author} {\bibinfo {author} {\bibfnamefont {C.-X.}\ \bibnamefont
  {Liu}}, \bibinfo {author} {\bibfnamefont {R.-X.}\ \bibnamefont {Zhang}}, \
  and\ \bibinfo {author} {\bibfnamefont {B.~K.}\ \bibnamefont {VanLeeuwen}},\
  }\href {\doibase 10.1103/PhysRevB.90.085304} {\bibfield  {journal} {\bibinfo
  {journal} {Phys. Rev. B}\ }\textbf {\bibinfo {volume} {90}},\ \bibinfo
  {pages} {085304} (\bibinfo {year} {2014})}\BibitemShut {NoStop}%
\bibitem [{\citenamefont {Alexandradinata}\ \emph
  {et~al.}(2014{\natexlab{a}})\citenamefont {Alexandradinata}, \citenamefont
  {Fang}, \citenamefont {Gilbert},\ and\ \citenamefont
  {Bernevig}}]{Alexandradinata2014g}%
  \BibitemOpen
  \bibfield  {author} {\bibinfo {author} {\bibfnamefont {A.}~\bibnamefont
  {Alexandradinata}}, \bibinfo {author} {\bibfnamefont {C.}~\bibnamefont
  {Fang}}, \bibinfo {author} {\bibfnamefont {M.~J.}\ \bibnamefont {Gilbert}}, \
  and\ \bibinfo {author} {\bibfnamefont {B.~A.}\ \bibnamefont {Bernevig}},\
  }\href {\doibase 10.1103/PhysRevLett.113.116403} {\bibfield  {journal}
  {\bibinfo  {journal} {Phys. Rev. Lett.}\ }\textbf {\bibinfo {volume} {113}},\
  \bibinfo {pages} {116403} (\bibinfo {year} {2014}{\natexlab{a}})}\BibitemShut
  {NoStop}%
\bibitem [{\citenamefont {Alexandradinata}\ \emph
  {et~al.}(2014{\natexlab{b}})\citenamefont {Alexandradinata}, \citenamefont
  {Dai},\ and\ \citenamefont {Bernevig}}]{Alexandradinata2014c}%
  \BibitemOpen
  \bibfield  {author} {\bibinfo {author} {\bibfnamefont {A.}~\bibnamefont
  {Alexandradinata}}, \bibinfo {author} {\bibfnamefont {X.}~\bibnamefont
  {Dai}}, \ and\ \bibinfo {author} {\bibfnamefont {B.~A.}\ \bibnamefont
  {Bernevig}},\ }\href {\doibase 10.1103/PhysRevB.89.155114} {\bibfield
  {journal} {\bibinfo  {journal} {Phys. Rev. B}\ }\textbf {\bibinfo {volume}
  {89}},\ \bibinfo {pages} {155114} (\bibinfo {year}
  {2014}{\natexlab{b}})}\BibitemShut {NoStop}%
\bibitem [{\citenamefont {Shiozaki}\ and\ \citenamefont
  {Sato}(2014)}]{Shiozaki2014}%
  \BibitemOpen
  \bibfield  {author} {\bibinfo {author} {\bibfnamefont {K.}~\bibnamefont
  {Shiozaki}}\ and\ \bibinfo {author} {\bibfnamefont {M.}~\bibnamefont
  {Sato}},\ }\href {\doibase 10.1103/PhysRevB.90.165114} {\bibfield  {journal}
  {\bibinfo  {journal} {Phys. Rev. B}\ }\textbf {\bibinfo {volume} {90}},\
  \bibinfo {pages} {165114} (\bibinfo {year} {2014})}\BibitemShut {NoStop}%
\bibitem [{\citenamefont {Fang}\ and\ \citenamefont
  {Fu}(2015{\natexlab{a}})}]{Fang2015}%
  \BibitemOpen
  \bibfield  {author} {\bibinfo {author} {\bibfnamefont {C.}~\bibnamefont
  {Fang}}\ and\ \bibinfo {author} {\bibfnamefont {L.}~\bibnamefont {Fu}},\
  }\href {\doibase 10.1103/PhysRevB.91.161105} {\bibfield  {journal} {\bibinfo
  {journal} {Phys. Rev. B}\ }\textbf {\bibinfo {volume} {91}},\ \bibinfo
  {pages} {161105} (\bibinfo {year} {2015}{\natexlab{a}})}\BibitemShut
  {NoStop}%
\bibitem [{\citenamefont {Shiozaki}\ \emph {et~al.}(2015)\citenamefont
  {Shiozaki}, \citenamefont {Sato},\ and\ \citenamefont {Gomi}}]{Shiozaki2015}%
  \BibitemOpen
  \bibfield  {author} {\bibinfo {author} {\bibfnamefont {K.}~\bibnamefont
  {Shiozaki}}, \bibinfo {author} {\bibfnamefont {M.}~\bibnamefont {Sato}}, \
  and\ \bibinfo {author} {\bibfnamefont {K.}~\bibnamefont {Gomi}},\ }\href
  {\doibase 10.1103/PhysRevB.91.155120} {\bibfield  {journal} {\bibinfo
  {journal} {Phys. Rev. B}\ }\textbf {\bibinfo {volume} {91}},\ \bibinfo
  {pages} {155120} (\bibinfo {year} {2015})}\BibitemShut {NoStop}%
\bibitem [{\citenamefont {Shiozaki}\ \emph {et~al.}(2016)\citenamefont
  {Shiozaki}, \citenamefont {Sato},\ and\ \citenamefont {Gomi}}]{Shiozaki2016}%
  \BibitemOpen
  \bibfield  {author} {\bibinfo {author} {\bibfnamefont {K.}~\bibnamefont
  {Shiozaki}}, \bibinfo {author} {\bibfnamefont {M.}~\bibnamefont {Sato}}, \
  and\ \bibinfo {author} {\bibfnamefont {K.}~\bibnamefont {Gomi}},\ }\href
  {\doibase 10.1103/PhysRevB.93.195413} {\bibfield  {journal} {\bibinfo
  {journal} {Phys. Rev. B}\ }\textbf {\bibinfo {volume} {93}},\ \bibinfo
  {pages} {195413} (\bibinfo {year} {2016})}\BibitemShut {NoStop}%
\bibitem [{\citenamefont {Shiozaki}\ \emph {et~al.}(2017)\citenamefont
  {Shiozaki}, \citenamefont {Sato},\ and\ \citenamefont {Gomi}}]{Shiozaki2017}%
  \BibitemOpen
  \bibfield  {author} {\bibinfo {author} {\bibfnamefont {K.}~\bibnamefont
  {Shiozaki}}, \bibinfo {author} {\bibfnamefont {M.}~\bibnamefont {Sato}}, \
  and\ \bibinfo {author} {\bibfnamefont {K.}~\bibnamefont {Gomi}},\ }\href
  {\doibase 10.1103/PhysRevB.95.235425} {\bibfield  {journal} {\bibinfo
  {journal} {Phys. Rev. B}\ }\textbf {\bibinfo {volume} {95}},\ \bibinfo
  {pages} {235425} (\bibinfo {year} {2017})}\BibitemShut {NoStop}%
\bibitem [{\citenamefont {Chiu}\ \emph {et~al.}(2013)\citenamefont {Chiu},
  \citenamefont {Yao},\ and\ \citenamefont {Ryu}}]{Classification_Chiu}%
  \BibitemOpen
  \bibfield  {author} {\bibinfo {author} {\bibfnamefont {C.-K.}\ \bibnamefont
  {Chiu}}, \bibinfo {author} {\bibfnamefont {H.}~\bibnamefont {Yao}}, \ and\
  \bibinfo {author} {\bibfnamefont {S.}~\bibnamefont {Ryu}},\ }\href@noop {}
  {\bibfield  {journal} {\bibinfo  {journal} {Phys. Rev. B}\ }\textbf {\bibinfo
  {volume} {88}},\ \bibinfo {pages} {075142} (\bibinfo {year}
  {2013})}\BibitemShut {NoStop}%
\bibitem [{\citenamefont {Morimoto}\ and\ \citenamefont
  {Furusaki}(2013)}]{AZ_mirror}%
  \BibitemOpen
  \bibfield  {author} {\bibinfo {author} {\bibfnamefont {T.}~\bibnamefont
  {Morimoto}}\ and\ \bibinfo {author} {\bibfnamefont {A.}~\bibnamefont
  {Furusaki}},\ }\href@noop {} {\bibfield  {journal} {\bibinfo  {journal}
  {Phys. Rev. B}\ }\textbf {\bibinfo {volume} {88}},\ \bibinfo {pages} {125129}
  (\bibinfo {year} {2013})}\BibitemShut {NoStop}%
\bibitem [{\citenamefont {Zak}(1979)}]{Zak1979}%
  \BibitemOpen
  \bibfield  {author} {\bibinfo {author} {\bibfnamefont {J.}~\bibnamefont
  {Zak}},\ }\href@noop {} {\bibfield  {journal} {\bibinfo  {journal} {Phys.
  Rev. B}\ }\textbf {\bibinfo {volume} {20}},\ \bibinfo {pages} {2228}
  (\bibinfo {year} {1979})}\BibitemShut {NoStop}%
\bibitem [{\citenamefont {Zak}(1981)}]{Zak1981}%
  \BibitemOpen
  \bibfield  {author} {\bibinfo {author} {\bibfnamefont {J.}~\bibnamefont
  {Zak}},\ }\href {http://link.aps.org/doi/10.1103/PhysRevB.23.2824} {\bibfield
   {journal} {\bibinfo  {journal} {Phys. Rev. B}\ }\textbf {\bibinfo {volume}
  {23}},\ \bibinfo {pages} {2824} (\bibinfo {year} {1981})}\BibitemShut
  {NoStop}%
\bibitem [{\citenamefont {Evarestov}\ and\ \citenamefont
  {Smirnov}(1984)}]{Evarestov1984}%
  \BibitemOpen
  \bibfield  {author} {\bibinfo {author} {\bibfnamefont {R.~A.}\ \bibnamefont
  {Evarestov}}\ and\ \bibinfo {author} {\bibfnamefont {V.~P.}\ \bibnamefont
  {Smirnov}},\ }\href@noop {} {\bibfield  {journal} {\bibinfo  {journal} {Phys.
  Stat. Sol}\ }\textbf {\bibinfo {volume} {122}},\ \bibinfo {pages} {231}
  (\bibinfo {year} {1984})}\BibitemShut {NoStop}%
\bibitem [{\citenamefont {Bacry}\ \emph {et~al.}(1988)\citenamefont {Bacry},
  \citenamefont {Michel},\ and\ \citenamefont {Zak}}]{Bacry1988}%
  \BibitemOpen
  \bibfield  {author} {\bibinfo {author} {\bibfnamefont {H.}~\bibnamefont
  {Bacry}}, \bibinfo {author} {\bibfnamefont {L.}~\bibnamefont {Michel}}, \
  and\ \bibinfo {author} {\bibfnamefont {J.}~\bibnamefont {Zak}},\ }\href
  {http://dx.doi.org/10.1007/BFb0012290} {\bibfield  {journal} {\bibinfo
  {journal} {Gr. Theor. Methods Phys.}\ }\textbf {\bibinfo {volume} {313}},\
  \bibinfo {pages} {289} (\bibinfo {year} {1988})}\BibitemShut {NoStop}%
\bibitem [{Note2()}]{Note2}%
  \BibitemOpen
  \bibinfo {note} {A non-simple band necessarily has rank $N{>}1$, but we do
  not require that a $(N{>}1)$-band subspace is connected by band touchings.
  Our definition of unlocalizable and localizable topological bands [cf.\ Fig.\
  \ref {fig:pie}.(a-b)] should in principle coincide with that of
  `topologically non-trivial bands' in Ref.\ \protect \rev@citealpnum
  {Bradlyn2017}, which `cannot be continued to any atomic limit without either
  closing a gap or breaking a symmetry.' Note that they have not defined
  topological triviality in the category of complex vector bundles, as we have
  done in this work. To verify their notion of non-triviality, it is necessary
  to construct a continuous family of Hamiltonians that symmetrically
  interpolates between the Hamiltonian of interest to an atomic-limit
  Hamiltonian (which is not uniquely defined). In comparison, our definition
  offers a diagnostic test (for the non-existence of locally-symmetric Wannier
  functions) that can be implemented on a single Hamiltonian of
  interest.}\BibitemShut {Stop}%
\bibitem [{Note3()}]{Note3}%
  \BibitemOpen
  \bibinfo {note} {For space groups with at least one nonsymmorphic element,
  bands must be nontrivially connected as a graph due to the monodromy of
  symmetry representations, as proven in the supplementary material of Ref.\
  \protect \rev@citealpnum {Holler2017}. Less general proofs exist for solids
  without spin-orbit coupling in Ref.\ \protect \rev@citealpnum
  {Parameswaran2013} and for band representations only in Ref.\ \protect
  \rev@citealpnum {Michel1999}. All type I (without time-reversal symmetry) and
  II (with time-reversal symmetry by itself) nonsymmorphic magnetic space
  groups have nonsymmorphic elements in $d{=}2$, but not in $d{\geq }3$. For
  nonsymmorphic space groups without nonsymmorphic elements, we are not aware
  of a general proof that simple bands do not exist; however this seems
  empirically to be true for specific case studies in $d{=}3$.\cite
  {Michel1999,Parameswaran2013,Po2016,Watanabe2017a}}\BibitemShut {NoStop}%
\bibitem [{\citenamefont {Freed}\ and\ \citenamefont
  {Moore}(2013)}]{Freed2013}%
  \BibitemOpen
  \bibfield  {author} {\bibinfo {author} {\bibfnamefont {D.~S.}\ \bibnamefont
  {Freed}}\ and\ \bibinfo {author} {\bibfnamefont {G.~W.}\ \bibnamefont
  {Moore}},\ }\href {\doibase 10.1007/s00023-013-0236-x} {\bibfield  {journal}
  {\bibinfo  {journal} {Ann. Henri Poincare}\ }\textbf {\bibinfo {volume}
  {14}},\ \bibinfo {pages} {1927} (\bibinfo {year} {2013})}\BibitemShut
  {NoStop}%
\bibitem [{\citenamefont {Po}\ \emph {et~al.}(2017)\citenamefont {Po},
  \citenamefont {Watanabe},\ and\ \citenamefont {Vishwanath}}]{Po}%
  \BibitemOpen
  \bibfield  {author} {\bibinfo {author} {\bibfnamefont {H.~C.}\ \bibnamefont
  {Po}}, \bibinfo {author} {\bibfnamefont {H.}~\bibnamefont {Watanabe}}, \ and\
  \bibinfo {author} {\bibfnamefont {A.}~\bibnamefont {Vishwanath}},\ }\href
  {https://arxiv.org/pdf/1709.06551.pdf} {\bibfield  {journal} {\bibinfo
  {journal} {arXiv}\ } (\bibinfo {year} {2017})},\ \Eprint
  {http://arxiv.org/abs/1709.06551} {arXiv:1709.06551} \BibitemShut {NoStop}%
\bibitem [{Note4()}]{Note4}%
  \BibitemOpen
  \bibinfo {note} {`Fragility' is a symmetry-enriched generalization of the
  notion of `stable triviality' in bundle theory.\cite {Hatcher2009,Read2017}
  Nontriviality of the fragile topological phases may manifest as a spectral
  flow in the holonomy over noncontractible Brillouin-zone loops.\cite
  {Alexandradinata2014c,Cano2017a}}\BibitemShut {NoStop}%
\bibitem [{\citenamefont {{De Nittis}}\ and\ \citenamefont
  {Gomi}(2014)}]{DeNittis2014}%
  \BibitemOpen
  \bibfield  {author} {\bibinfo {author} {\bibfnamefont {G.}~\bibnamefont {{De
  Nittis}}}\ and\ \bibinfo {author} {\bibfnamefont {K.}~\bibnamefont {Gomi}},\
  }\href {\doibase 10.1016/J.GEOMPHYS.2014.07.036} {\bibfield  {journal}
  {\bibinfo  {journal} {J. Geom. Phys.}\ }\textbf {\bibinfo {volume} {86}},\
  \bibinfo {pages} {303} (\bibinfo {year} {2014})}\BibitemShut {NoStop}%
\bibitem [{\citenamefont {{De Nittis}}\ and\ \citenamefont
  {Gomi}(2017)}]{DeNittis2017}%
  \BibitemOpen
  \bibfield  {author} {\bibinfo {author} {\bibfnamefont {G.}~\bibnamefont {{De
  Nittis}}}\ and\ \bibinfo {author} {\bibfnamefont {K.}~\bibnamefont {Gomi}},\
  }\href {\doibase 10.1007/s11005-017-1029-9} {\bibfield  {journal} {\bibinfo
  {journal} {Lett. Math. Phys.}\ ,\ \bibinfo {pages} {1}} (\bibinfo {year}
  {2017})}\BibitemShut {NoStop}%
\bibitem [{\citenamefont {Thouless}\ \emph {et~al.}(1982)\citenamefont
  {Thouless}, \citenamefont {Kohmoto}, \citenamefont {Nightingale},\ and\
  \citenamefont {den Nijs}}]{Thouless1982}%
  \BibitemOpen
  \bibfield  {author} {\bibinfo {author} {\bibfnamefont {D.~J.}\ \bibnamefont
  {Thouless}}, \bibinfo {author} {\bibfnamefont {M.}~\bibnamefont {Kohmoto}},
  \bibinfo {author} {\bibfnamefont {M.~P.}\ \bibnamefont {Nightingale}}, \ and\
  \bibinfo {author} {\bibfnamefont {M.}~\bibnamefont {den Nijs}},\ }\href@noop
  {} {\bibfield  {journal} {\bibinfo  {journal} {Phys. Rev. Lett.}\ }\textbf
  {\bibinfo {volume} {49}},\ \bibinfo {pages} {405} (\bibinfo {year}
  {1982})}\BibitemShut {NoStop}%
\bibitem [{\citenamefont {Peterson}(1959)}]{Peterson1959}%
  \BibitemOpen
  \bibfield  {author} {\bibinfo {author} {\bibfnamefont {F.~P.}\ \bibnamefont
  {Peterson}},\ }\href {\doibase 10.2307/1970191} {\bibfield  {journal}
  {\bibinfo  {journal} {Ann. Math.}\ }\textbf {\bibinfo {volume} {69}},\
  \bibinfo {pages} {414} (\bibinfo {year} {1959})}\BibitemShut {NoStop}%
\bibitem [{\citenamefont {Husemoller}(1994)}]{Husemoller1994}%
  \BibitemOpen
  \bibfield  {author} {\bibinfo {author} {\bibfnamefont {D.}~\bibnamefont
  {Husemoller}},\ }\href {\doibase 10.1007/978-1-4757-2261-1} {\emph {\bibinfo
  {title} {{Fibre Bundles}}}},\ \bibinfo {series} {Graduate Texts in
  Mathematics}, Vol.~\bibinfo {volume} {20}\ (\bibinfo  {publisher}
  {Springer},\ \bibinfo {address} {New York},\ \bibinfo {year}
  {1994})\BibitemShut {NoStop}%
\bibitem [{Note5()}]{Note5}%
  \BibitemOpen
  \bibinfo {note} {Bloch functions do not have to correspond to energy bands;
  sometimes these are referred as quasi-Bloch functions.\cite
  {Nenciu1991,Brouder2007,Panati2007}}\BibitemShut {NoStop}%
\bibitem [{\citenamefont {Price}\ \emph {et~al.}(2015)\citenamefont {Price},
  \citenamefont {Zilberberg}, \citenamefont {Ozawa}, \citenamefont
  {Carusotto},\ and\ \citenamefont {Goldman}}]{Price2015}%
  \BibitemOpen
  \bibfield  {author} {\bibinfo {author} {\bibfnamefont {H.}~\bibnamefont
  {Price}}, \bibinfo {author} {\bibfnamefont {O.}~\bibnamefont {Zilberberg}},
  \bibinfo {author} {\bibfnamefont {T.}~\bibnamefont {Ozawa}}, \bibinfo
  {author} {\bibfnamefont {I.}~\bibnamefont {Carusotto}}, \ and\ \bibinfo
  {author} {\bibfnamefont {N.}~\bibnamefont {Goldman}},\ }\href {\doibase
  10.1103/PhysRevLett.115.195303} {\bibfield  {journal} {\bibinfo  {journal}
  {Phys. Rev. Lett.}\ }\textbf {\bibinfo {volume} {115}},\ \bibinfo {pages}
  {195303} (\bibinfo {year} {2015})}\BibitemShut {NoStop}%
\bibitem [{\citenamefont {Lohse}\ \emph {et~al.}(2018)\citenamefont {Lohse},
  \citenamefont {Schweizer}, \citenamefont {Price}, \citenamefont
  {Zilberberg},\ and\ \citenamefont {Bloch}}]{Lohse2018}%
  \BibitemOpen
  \bibfield  {author} {\bibinfo {author} {\bibfnamefont {M.}~\bibnamefont
  {Lohse}}, \bibinfo {author} {\bibfnamefont {C.}~\bibnamefont {Schweizer}},
  \bibinfo {author} {\bibfnamefont {H.~M.}\ \bibnamefont {Price}}, \bibinfo
  {author} {\bibfnamefont {O.}~\bibnamefont {Zilberberg}}, \ and\ \bibinfo
  {author} {\bibfnamefont {I.}~\bibnamefont {Bloch}},\ }\href {\doibase
  10.1038/nature25000} {\bibfield  {journal} {\bibinfo  {journal} {Nature}\
  }\textbf {\bibinfo {volume} {553}},\ \bibinfo {pages} {55} (\bibinfo {year}
  {2018})}\BibitemShut {NoStop}%
\bibitem [{\citenamefont {Ningyuan}\ \emph {et~al.}(2015)\citenamefont
  {Ningyuan}, \citenamefont {Owens}, \citenamefont {Sommer}, \citenamefont
  {Schuster},\ and\ \citenamefont {Simon}}]{Ningyuan2015}%
  \BibitemOpen
  \bibfield  {author} {\bibinfo {author} {\bibfnamefont {J.}~\bibnamefont
  {Ningyuan}}, \bibinfo {author} {\bibfnamefont {C.}~\bibnamefont {Owens}},
  \bibinfo {author} {\bibfnamefont {A.}~\bibnamefont {Sommer}}, \bibinfo
  {author} {\bibfnamefont {D.}~\bibnamefont {Schuster}}, \ and\ \bibinfo
  {author} {\bibfnamefont {J.}~\bibnamefont {Simon}},\ }\href {\doibase
  10.1103/PhysRevX.5.021031} {\bibfield  {journal} {\bibinfo  {journal} {Phys.
  Rev. X}\ }\textbf {\bibinfo {volume} {5}},\ \bibinfo {pages} {021031}
  (\bibinfo {year} {2015})}\BibitemShut {NoStop}%
\bibitem [{\citenamefont {Albert}\ \emph {et~al.}(2015)\citenamefont {Albert},
  \citenamefont {Glazman},\ and\ \citenamefont {Jiang}}]{Albert2015}%
  \BibitemOpen
  \bibfield  {author} {\bibinfo {author} {\bibfnamefont {V.~V.}\ \bibnamefont
  {Albert}}, \bibinfo {author} {\bibfnamefont {L.~I.}\ \bibnamefont {Glazman}},
  \ and\ \bibinfo {author} {\bibfnamefont {L.}~\bibnamefont {Jiang}},\ }\href
  {\doibase 10.1103/PhysRevLett.114.173902} {\bibfield  {journal} {\bibinfo
  {journal} {Phys. Rev. Lett.}\ }\textbf {\bibinfo {volume} {114}},\ \bibinfo
  {pages} {173902} (\bibinfo {year} {2015})}\BibitemShut {NoStop}%
\bibitem [{\citenamefont {Lee}\ \emph {et~al.}(2018)\citenamefont {Lee},
  \citenamefont {Li}, \citenamefont {Jin}, \citenamefont {Liu},\ and\
  \citenamefont {Zhang}}]{Lee2018}%
  \BibitemOpen
  \bibfield  {author} {\bibinfo {author} {\bibfnamefont {C.~H.}\ \bibnamefont
  {Lee}}, \bibinfo {author} {\bibfnamefont {G.}~\bibnamefont {Li}}, \bibinfo
  {author} {\bibfnamefont {G.}~\bibnamefont {Jin}}, \bibinfo {author}
  {\bibfnamefont {Y.}~\bibnamefont {Liu}}, \ and\ \bibinfo {author}
  {\bibfnamefont {X.}~\bibnamefont {Zhang}},\ }\href {\doibase
  10.1103/PhysRevB.97.085110} {\bibfield  {journal} {\bibinfo  {journal} {Phys.
  Rev. B}\ }\textbf {\bibinfo {volume} {97}},\ \bibinfo {pages} {085110}
  (\bibinfo {year} {2018})}\BibitemShut {NoStop}%
\bibitem [{\citenamefont {Hatcher}(2009)}]{Hatcher2009}%
  \BibitemOpen
  \bibfield  {author} {\bibinfo {author} {\bibfnamefont {A.}~\bibnamefont
  {Hatcher}},\ }\href@noop {} {\emph {\bibinfo {title} {{Vector Bundles and
  K-Theory}}}},\ \bibinfo {number} {May}\ (\bibinfo {year} {2009})\BibitemShut
  {NoStop}%
\bibitem [{Note6()}]{Note6}%
  \BibitemOpen
  \bibinfo {note} {Precisely, $e^{-i\protect \bm {k}\cdot \protect \bm
  {r}}H_0e^{i\protect \bm {k}\cdot \protect \bm {r}}$ should be in the entire
  analytic family of type A, as defined in Ref.\ \protect \rev@citealpnum
  {Reed}.}\BibitemShut {Stop}%
\bibitem [{Note7()}]{Note7}%
  \BibitemOpen
  \bibinfo {note} {Dimension-dependent conditions on $V$ are stated in Ref.\
  \protect \rev@citealpnum {Reed}, Theorem XIII.99: $V{\in }L^p($unit
  cell$,d^d\protect \bm {r})$ for $p{=}2$ if $d{\le }3$, $p{>}2$ if $d{=}4$ and
  $p{=}d/2$ if $d{\ge }5$.}\BibitemShut {Stop}%
\bibitem [{\citenamefont {Reed}\ and\ \citenamefont {Simon}(1978)}]{Reed}%
  \BibitemOpen
  \bibfield  {author} {\bibinfo {author} {\bibfnamefont {M.}~\bibnamefont
  {Reed}}\ and\ \bibinfo {author} {\bibfnamefont {B.}~\bibnamefont {Simon}},\
  }\href@noop {} {\emph {\bibinfo {title} {{Methods of modern mathematical
  physics. IV, Analysis of operators}}}}\ (\bibinfo  {publisher} {Academic
  Press},\ \bibinfo {year} {1978})\BibitemShut {NoStop}%
\bibitem [{\citenamefont {Nenciu}(1991)}]{Nenciu1991}%
  \BibitemOpen
  \bibfield  {author} {\bibinfo {author} {\bibfnamefont {G.}~\bibnamefont
  {Nenciu}},\ }\href {\doibase 10.1103/RevModPhys.34.645} {\bibfield  {journal}
  {\bibinfo  {journal} {Rev. Mod. Phys.}\ }\textbf {\bibinfo {volume} {34}},\
  \bibinfo {pages} {645} (\bibinfo {year} {1991})}\BibitemShut {NoStop}%
\bibitem [{\citenamefont {Evarestov}\ and\ \citenamefont
  {Smirnov}(1997)}]{Evarestov}%
  \BibitemOpen
  \bibfield  {author} {\bibinfo {author} {\bibfnamefont {R.~A.}\ \bibnamefont
  {Evarestov}}\ and\ \bibinfo {author} {\bibfnamefont {V.~P.}\ \bibnamefont
  {Smirnov}},\ }\href@noop {} {\emph {\bibinfo {title} {{Site Symmetry in
  Crystals}}}},\ Vol.~\bibinfo {volume} {1}\ (\bibinfo  {publisher}
  {Springer},\ \bibinfo {address} {Berlin, Heidelberg},\ \bibinfo {year}
  {1997})\ pp.\ \bibinfo {pages} {1689--1699}\BibitemShut {NoStop}%
\bibitem [{Note8()}]{Note8}%
  \BibitemOpen
  \bibinfo {note} {This action of $g$ on $f$ defines a regular
  representation\cite {Bacry1993} which is known to be linear.}\BibitemShut
  {Stop}%
\bibitem [{\citenamefont {Lowdin}(1950)}]{Lodwin1950}%
  \BibitemOpen
  \bibfield  {author} {\bibinfo {author} {\bibfnamefont {P.}~\bibnamefont
  {Lowdin}},\ }\href@noop {} {\bibfield  {journal} {\bibinfo  {journal} {J.
  Chem. Phys.}\ }\textbf {\bibinfo {volume} {18}},\ \bibinfo {pages} {365}
  (\bibinfo {year} {1950})}\BibitemShut {NoStop}%
\bibitem [{Note9()}]{Note9}%
  \BibitemOpen
  \bibinfo {note} {This follows from analytic perturbation theory, e.g., see
  Ref.\ \protect \rev@citealpnum {Dubail2015} for the tight-binding $H_0$, and
  the Kato-Rellich theorem in Ref.\ \protect \rev@citealpnum {Reed} for the
  Schroedinger $H_0$.}\BibitemShut {Stop}%
\bibitem [{Note10()}]{Note10}%
  \BibitemOpen
  \bibinfo {note} {A complex neighborhood of $T^d$ may be identified as a
  domain of holomorphy in $\protect \mathbb {C}^d$ and therefore a Stein space.
  Solving the second Cousin problem over a Stein space is equivalent to proving
  the existence of a global analytic section for a topologically trivial line
  bundle; this has been carried out in Ref.\ \protect \rev@citealpnum
  {Hormander1989}. See also related discussions falling under the `Grauert-Oka
  principle'\cite {Oka1939,Grauert1958} in Ref.\ \protect \rev@citealpnum
  {Panati2007,Huckleberry2013}. In fact, the non-abelian second Cousin problem
  has also been solved using sheaf theory;\cite
  {Grauert1958,Gindikin1986,Henkin1998} this implies, for a topologically
  trivial band of rank $N{\geq }1$, that there exists analytic and periodic
  Bloch functions which span the $N$-dimensional vector space at each $\protect
  \bm {k}$.}\BibitemShut {Stop}%
\bibitem [{\citenamefont {Michel}\ and\ \citenamefont
  {Zak}(1999)}]{Michel1999}%
  \BibitemOpen
  \bibfield  {author} {\bibinfo {author} {\bibfnamefont {L.}~\bibnamefont
  {Michel}}\ and\ \bibinfo {author} {\bibfnamefont {J.}~\bibnamefont {Zak}},\
  }\href {\doibase 10.1103/PhysRevB.59.5998} {\bibfield  {journal} {\bibinfo
  {journal} {Phys. Rev. B}\ }\textbf {\bibinfo {volume} {59}},\ \bibinfo
  {pages} {5998} (\bibinfo {year} {1999})}\BibitemShut {NoStop}%
\bibitem [{\citenamefont {King-Smith}\ and\ \citenamefont
  {Vanderbilt}(1993)}]{King-Smith1993}%
  \BibitemOpen
  \bibfield  {author} {\bibinfo {author} {\bibfnamefont {R.~D.}\ \bibnamefont
  {King-Smith}}\ and\ \bibinfo {author} {\bibfnamefont {D.}~\bibnamefont
  {Vanderbilt}},\ }\href {\doibase 10.1103/PhysRevB.47.1651} {\bibfield
  {journal} {\bibinfo  {journal} {Rapid Commun.}\ }\textbf {\bibinfo {volume}
  {47}},\ \bibinfo {pages} {1651} (\bibinfo {year} {1993})}\BibitemShut
  {NoStop}%
\bibitem [{\citenamefont {Vanderbilt}\ and\ \citenamefont
  {King-Smith}(1993)}]{Vanderbilt1993}%
  \BibitemOpen
  \bibfield  {author} {\bibinfo {author} {\bibfnamefont {D.}~\bibnamefont
  {Vanderbilt}}\ and\ \bibinfo {author} {\bibfnamefont {R.~D.}\ \bibnamefont
  {King-Smith}},\ }\href {\doibase 10.1103/PhysRevB.48.4442} {\bibfield
  {journal} {\bibinfo  {journal} {Phys. Rev. B}\ }\textbf {\bibinfo {volume}
  {48}},\ \bibinfo {pages} {4442} (\bibinfo {year} {1993})}\BibitemShut
  {NoStop}%
\bibitem [{Note11()}]{Note11}%
  \BibitemOpen
  \bibinfo {note} {Generally, one constructs Wannier functions for all Wyckoff
  positions in the orbit $G{\circ }\protect \bm {\varpi }$; for
  unit-multiplicity Wyckoff positions, these Wannier functions are just the
  $BL$-translates of the single Wannier function.}\BibitemShut {Stop}%
\bibitem [{Note12()}]{Note12}%
  \BibitemOpen
  \bibinfo {note} {This follows because $[T^d,U(1)]$ are in one-to-one
  correspondence with cohomology classes in $H^1(T^d;\protect \mathbb {Z})$.
  Applying the Universal Coefficient Theorem, $H^1(T^d;\protect \mathbb
  {Z}){\protect \cong }\protect \mathbb {Z}^d$.}\BibitemShut {Stop}%
\bibitem [{Note13()}]{Note13}%
  \BibitemOpen
  \bibinfo {note} {For any two representations ${\setbox \z@ \hbox
  {\frozen@everymath \@emptytoks \mathsurround \z@ $\nulldelimiterspace \z@
  \left (\vcenter to\@ne \big@size {}\right .$}\box \z@ } \rho _p^0(\protect
  \bm {k}),\rho _p^1(\protect \bm {k}) {\setbox \z@ \hbox {\frozen@everymath
  \@emptytoks \mathsurround \z@ $\nulldelimiterspace \z@ \left )\vcenter to\@ne
  \big@size {}\right .$}\box \z@ }$ of $G$ that are related by a gauge
  transformation with periodic $\phi (\protect \bm {k})$, there exists a
  continuous interpolation $\rho _p^s(\protect \bm {k})$ ($s\in [0,1]$) that is
  itself analytic and periodic, and a representation of $G$ throughout the
  interpolation.}\BibitemShut {Stop}%
\bibitem [{\citenamefont {Lax}(1974)}]{MelvinLax1974}%
  \BibitemOpen
  \bibfield  {author} {\bibinfo {author} {\bibfnamefont {M.}~\bibnamefont
  {Lax}},\ }\href
  {https://books.google.com/books?id=IZkoAwAAQBAJ{\&}pg=PA455{\&}lpg=PA455{\&}dq={\%}5B6{\%}5D+M.+Lax,+Symmetry+principles+in+solid+state+and+molecular+physics.{\&}source=bl{\&}ots=c2XEYWzYW1{\&}sig=uuydHCp2fqXvrWPIMNiOlfkpdIE{\&}hl=en{\&}sa=X{\&}ved=0ahUKEwj05Ofcg8{\_}UAhVD2T4KHV4hDRoQ6AEIOTA}
  {\emph {\bibinfo {title} {{Symmetry Principles in Solid State and Molecular
  Physics}}}}\ (\bibinfo  {publisher} {J. Wiley},\ \bibinfo {address} {New
  York},\ \bibinfo {year} {1974})\BibitemShut {NoStop}%
\bibitem [{\citenamefont {Tinkham}(2003)}]{Tinkham2003}%
  \BibitemOpen
  \bibfield  {author} {\bibinfo {author} {\bibfnamefont {M.}~\bibnamefont
  {Tinkham}},\ }\href@noop {} {\emph {\bibinfo {title} {{Group theory and
  quantum mechanics}}}}\ (\bibinfo  {publisher} {Dover Publications},\ \bibinfo
  {address} {New York},\ \bibinfo {year} {2003})\BibitemShut {NoStop}%
\bibitem [{Note14()}]{Note14}%
  \BibitemOpen
  \bibinfo {note} {Precisely, we mean a Whitney sum of two vector bundles,
  where at each $\protect \bm {k}$ the two vector spaces are directly
  summed.}\BibitemShut {Stop}%
\bibitem [{\citenamefont {Kitaev}(2009)}]{kitaev_periodictable}%
  \BibitemOpen
  \bibfield  {author} {\bibinfo {author} {\bibfnamefont {A.}~\bibnamefont
  {Kitaev}},\ }\href@noop {} {\bibfield  {journal} {\bibinfo  {journal} {AIP
  Conf. Proc.}\ }\textbf {\bibinfo {volume} {1134}},\ \bibinfo {pages} {22}
  (\bibinfo {year} {2009})}\BibitemShut {NoStop}%
\bibitem [{\citenamefont {Bacry}(1993)}]{Bacry1993}%
  \BibitemOpen
  \bibfield  {author} {\bibinfo {author} {\bibfnamefont {H.}~\bibnamefont
  {Bacry}},\ }\href {\doibase 10.1007/BF02096648} {\bibfield  {journal}
  {\bibinfo  {journal} {Commun. Math. Phys.}\ }\textbf {\bibinfo {volume}
  {153}},\ \bibinfo {pages} {359} (\bibinfo {year} {1993})}\BibitemShut
  {NoStop}%
\bibitem [{\citenamefont {Bilbao}()}]{bilbao}%
  \BibitemOpen
  \bibfield  {author} {\bibinfo {author} {\bibnamefont {Bilbao}},\ }\href
  {http://www.cryst.ehu.es/} {\enquote {\bibinfo {title}
  {{http://www.cryst.ehu.es/}},}\ }\BibitemShut {NoStop}%
\bibitem [{\citenamefont {Hughes}\ \emph {et~al.}(2011)\citenamefont {Hughes},
  \citenamefont {Prodan},\ and\ \citenamefont {Bernevig}}]{Hughes2011a}%
  \BibitemOpen
  \bibfield  {author} {\bibinfo {author} {\bibfnamefont {T.~L.}\ \bibnamefont
  {Hughes}}, \bibinfo {author} {\bibfnamefont {E.}~\bibnamefont {Prodan}}, \
  and\ \bibinfo {author} {\bibfnamefont {B.~A.}\ \bibnamefont {Bernevig}},\
  }\href {\doibase 10.1103/PhysRevB.83.245132} {\bibfield  {journal} {\bibinfo
  {journal} {Phys. Rev. B}\ }\textbf {\bibinfo {volume} {83}},\ \bibinfo
  {pages} {245132} (\bibinfo {year} {2011})}\BibitemShut {NoStop}%
\bibitem [{\citenamefont {Turner}\ \emph {et~al.}(2012)\citenamefont {Turner},
  \citenamefont {Zhang}, \citenamefont {Mong},\ and\ \citenamefont
  {Vishwanath}}]{Turner2012a}%
  \BibitemOpen
  \bibfield  {author} {\bibinfo {author} {\bibfnamefont {A.~M.}\ \bibnamefont
  {Turner}}, \bibinfo {author} {\bibfnamefont {Y.}~\bibnamefont {Zhang}},
  \bibinfo {author} {\bibfnamefont {R.~S.~K.}\ \bibnamefont {Mong}}, \ and\
  \bibinfo {author} {\bibfnamefont {A.}~\bibnamefont {Vishwanath}},\ }\href
  {\doibase 10.1103/PhysRevB.85.165120} {\bibfield  {journal} {\bibinfo
  {journal} {Phys. Rev. B}\ }\textbf {\bibinfo {volume} {85}},\ \bibinfo
  {pages} {165120} (\bibinfo {year} {2012})}\BibitemShut {NoStop}%
\bibitem [{\citenamefont {Fang}\ \emph {et~al.}(2012)\citenamefont {Fang},
  \citenamefont {Gilbert},\ and\ \citenamefont {Bernevig}}]{Fang2012}%
  \BibitemOpen
  \bibfield  {author} {\bibinfo {author} {\bibfnamefont {C.}~\bibnamefont
  {Fang}}, \bibinfo {author} {\bibfnamefont {M.~J.}\ \bibnamefont {Gilbert}}, \
  and\ \bibinfo {author} {\bibfnamefont {B.~A.}\ \bibnamefont {Bernevig}},\
  }\href {\doibase 10.1103/PhysRevB.86.115112} {\bibfield  {journal} {\bibinfo
  {journal} {Phys. Rev. B}\ }\textbf {\bibinfo {volume} {86}},\ \bibinfo
  {pages} {115112} (\bibinfo {year} {2012})}\BibitemShut {NoStop}%
\bibitem [{\citenamefont {Song}\ \emph {et~al.}(2017)\citenamefont {Song},
  \citenamefont {Zhang}, \citenamefont {Fang},\ and\ \citenamefont
  {Fang}}]{Song2017}%
  \BibitemOpen
  \bibfield  {author} {\bibinfo {author} {\bibfnamefont {Z.}~\bibnamefont
  {Song}}, \bibinfo {author} {\bibfnamefont {T.}~\bibnamefont {Zhang}},
  \bibinfo {author} {\bibfnamefont {Z.}~\bibnamefont {Fang}}, \ and\ \bibinfo
  {author} {\bibfnamefont {C.}~\bibnamefont {Fang}},\ }\href
  {http://arxiv.org/abs/1711.11049} {\  (\bibinfo {year} {2017})},\ \Eprint
  {http://arxiv.org/abs/1711.11049} {arXiv:1711.11049} \BibitemShut {NoStop}%
\bibitem [{Note15()}]{Note15}%
  \BibitemOpen
  \bibinfo {note} {It is characterized by a unit relative winding number -- a
  topological invariant first discovered in Ref.\ \protect \rev@citealpnum
  {Alexandradinata2014c}. This invariant has also been used to characterize
  other fragile phases.\cite {Cano2017a}}\BibitemShut {NoStop}%
\bibitem [{\citenamefont {Parameswaran}\ \emph {et~al.}(2013)\citenamefont
  {Parameswaran}, \citenamefont {Turner}, \citenamefont {Arovas},\ and\
  \citenamefont {Vishwanath}}]{Parameswaran2013}%
  \BibitemOpen
  \bibfield  {author} {\bibinfo {author} {\bibfnamefont {S.~A.}\ \bibnamefont
  {Parameswaran}}, \bibinfo {author} {\bibfnamefont {A.~M.}\ \bibnamefont
  {Turner}}, \bibinfo {author} {\bibfnamefont {D.~P.}\ \bibnamefont {Arovas}},
  \ and\ \bibinfo {author} {\bibfnamefont {A.}~\bibnamefont {Vishwanath}},\
  }\href {\doibase 10.1038/nphys2600} {\bibfield  {journal} {\bibinfo
  {journal} {Nat. Phys.}\ }\textbf {\bibinfo {volume} {9}},\ \bibinfo {pages}
  {299} (\bibinfo {year} {2013})}\BibitemShut {NoStop}%
\bibitem [{\citenamefont {Po}\ \emph {et~al.}(2016)\citenamefont {Po},
  \citenamefont {Watanabe}, \citenamefont {Zaletel},\ and\ \citenamefont
  {Vishwanath}}]{Po2016}%
  \BibitemOpen
  \bibfield  {author} {\bibinfo {author} {\bibfnamefont {H.~C.}\ \bibnamefont
  {Po}}, \bibinfo {author} {\bibfnamefont {H.}~\bibnamefont {Watanabe}},
  \bibinfo {author} {\bibfnamefont {M.~P.}\ \bibnamefont {Zaletel}}, \ and\
  \bibinfo {author} {\bibfnamefont {A.}~\bibnamefont {Vishwanath}},\ }\href
  {\doibase 10.1126/sciadv.1501782} {\bibfield  {journal} {\bibinfo  {journal}
  {Sci. Adv.}\ }\textbf {\bibinfo {volume} {2}},\ \bibinfo {pages} {e1501782}
  (\bibinfo {year} {2016})}\BibitemShut {NoStop}%
\bibitem [{\citenamefont {Bouckaert}\ \emph {et~al.}(1936)\citenamefont
  {Bouckaert}, \citenamefont {Smoluchowski},\ and\ \citenamefont
  {Wigner}}]{Bouckaert1936}%
  \BibitemOpen
  \bibfield  {author} {\bibinfo {author} {\bibfnamefont {L.~P.}\ \bibnamefont
  {Bouckaert}}, \bibinfo {author} {\bibfnamefont {R.}~\bibnamefont
  {Smoluchowski}}, \ and\ \bibinfo {author} {\bibfnamefont {E.}~\bibnamefont
  {Wigner}},\ }\href {\doibase 10.1103/PhysRev.50.58} {\bibfield  {journal}
  {\bibinfo  {journal} {Phys. Rev.}\ }\textbf {\bibinfo {volume} {50}},\
  \bibinfo {pages} {58} (\bibinfo {year} {1936})}\BibitemShut {NoStop}%
\bibitem [{\citenamefont {Herring}(1937)}]{Herring1937}%
  \BibitemOpen
  \bibfield  {author} {\bibinfo {author} {\bibfnamefont {C.}~\bibnamefont
  {Herring}},\ }\href {\doibase 10.1103/PhysRev.52.361} {\bibfield  {journal}
  {\bibinfo  {journal} {Phys. Rev.}\ }\textbf {\bibinfo {volume} {52}},\
  \bibinfo {pages} {361} (\bibinfo {year} {1937})}\BibitemShut {NoStop}%
\bibitem [{\citenamefont {Watanabe}\ \emph {et~al.}(2016)\citenamefont
  {Watanabe}, \citenamefont {Po}, \citenamefont {Zaletel},\ and\ \citenamefont
  {Vishwanath}}]{Watanabe2016}%
  \BibitemOpen
  \bibfield  {author} {\bibinfo {author} {\bibfnamefont {H.}~\bibnamefont
  {Watanabe}}, \bibinfo {author} {\bibfnamefont {H.~C.}\ \bibnamefont {Po}},
  \bibinfo {author} {\bibfnamefont {M.~P.}\ \bibnamefont {Zaletel}}, \ and\
  \bibinfo {author} {\bibfnamefont {A.}~\bibnamefont {Vishwanath}},\ }\href
  {\doibase 10.1103/PhysRevLett.117.096404} {\bibfield  {journal} {\bibinfo
  {journal} {Phys. Rev. Lett.}\ }\textbf {\bibinfo {volume} {117}},\ \bibinfo
  {pages} {096404} (\bibinfo {year} {2016})}\BibitemShut {NoStop}%
\bibitem [{\citenamefont {Lu}\ \emph {et~al.}(2014)\citenamefont {Lu},
  \citenamefont {Joannopoulos},\ and\ \citenamefont
  {Solja{\v{c}}i{\'{c}}}}]{Lu2014}%
  \BibitemOpen
  \bibfield  {author} {\bibinfo {author} {\bibfnamefont {L.}~\bibnamefont
  {Lu}}, \bibinfo {author} {\bibfnamefont {J.~D.}\ \bibnamefont
  {Joannopoulos}}, \ and\ \bibinfo {author} {\bibfnamefont {M.}~\bibnamefont
  {Solja{\v{c}}i{\'{c}}}},\ }\href {\doibase 10.1038/nphoton.2014.248}
  {\bibfield  {journal} {\bibinfo  {journal} {Nat. Photonics}\ }\textbf
  {\bibinfo {volume} {8}},\ \bibinfo {pages} {821} (\bibinfo {year}
  {2014})}\BibitemShut {NoStop}%
\bibitem [{\citenamefont {Lu}\ \emph {et~al.}(2016)\citenamefont {Lu},
  \citenamefont {Fang}, \citenamefont {Fu}, \citenamefont {Johnson},
  \citenamefont {Joannopoulos},\ and\ \citenamefont
  {Solja{\v{c}}i{\'{c}}}}]{Lu2016}%
  \BibitemOpen
  \bibfield  {author} {\bibinfo {author} {\bibfnamefont {L.}~\bibnamefont
  {Lu}}, \bibinfo {author} {\bibfnamefont {C.}~\bibnamefont {Fang}}, \bibinfo
  {author} {\bibfnamefont {L.}~\bibnamefont {Fu}}, \bibinfo {author}
  {\bibfnamefont {S.~G.}\ \bibnamefont {Johnson}}, \bibinfo {author}
  {\bibfnamefont {J.~D.}\ \bibnamefont {Joannopoulos}}, \ and\ \bibinfo
  {author} {\bibfnamefont {M.}~\bibnamefont {Solja{\v{c}}i{\'{c}}}},\ }\href
  {\doibase 10.1038/nphys3611} {\bibfield  {journal} {\bibinfo  {journal} {Nat.
  Phys.}\ }\textbf {\bibinfo {volume} {12}},\ \bibinfo {pages} {337} (\bibinfo
  {year} {2016})}\BibitemShut {NoStop}%
\bibitem [{\citenamefont {S{\"{u}}sstrunk}\ and\ \citenamefont
  {Huber}(2015)}]{Susstrunk2015}%
  \BibitemOpen
  \bibfield  {author} {\bibinfo {author} {\bibfnamefont {R.}~\bibnamefont
  {S{\"{u}}sstrunk}}\ and\ \bibinfo {author} {\bibfnamefont {S.~D.}\
  \bibnamefont {Huber}},\ }\href {\doibase 10.1126/science.aab0239} {\bibfield
  {journal} {\bibinfo  {journal} {Science}\ }\textbf {\bibinfo {volume}
  {349}},\ \bibinfo {pages} {47} (\bibinfo {year} {2015})}\BibitemShut
  {NoStop}%
\bibitem [{\citenamefont {S{\"{u}}sstrunk}\ and\ \citenamefont
  {Huber}(2016)}]{Susstrunk2016}%
  \BibitemOpen
  \bibfield  {author} {\bibinfo {author} {\bibfnamefont {R.}~\bibnamefont
  {S{\"{u}}sstrunk}}\ and\ \bibinfo {author} {\bibfnamefont {S.~D.}\
  \bibnamefont {Huber}},\ }\href {\doibase 10.1073/pnas.1605462113} {\bibfield
  {journal} {\bibinfo  {journal} {Proc. Natl. Acad. Sci. U. S. A.}\ }\textbf
  {\bibinfo {volume} {113}},\ \bibinfo {pages} {E4767} (\bibinfo {year}
  {2016})}\BibitemShut {NoStop}%
\bibitem [{\citenamefont {Grauert}(1958)}]{Grauert1958}%
  \BibitemOpen
  \bibfield  {author} {\bibinfo {author} {\bibfnamefont {H.}~\bibnamefont
  {Grauert}},\ }\href {\doibase 10.1007/BF01351803} {\bibfield  {journal}
  {\bibinfo  {journal} {Math. Ann.}\ }\textbf {\bibinfo {volume} {135}},\
  \bibinfo {pages} {263} (\bibinfo {year} {1958})}\BibitemShut {NoStop}%
\bibitem [{\citenamefont {Gindikin}\ and\ \citenamefont
  {Khenkin}(1986)}]{Gindikin1986}%
  \BibitemOpen
  \bibfield  {author} {\bibinfo {author} {\bibfnamefont {S.~G.}\ \bibnamefont
  {Gindikin}}\ and\ \bibinfo {author} {\bibfnamefont {G.~M.}\ \bibnamefont
  {Khenkin}},\ }\href
  {https://link.springer.com/content/pdf/10.1007{\%}2F978-3-642-61263-3.pdf}
  {\emph {\bibinfo {title} {{Several complex variables IV}}}}\ (\bibinfo
  {publisher} {Springer, Berlin, Heidelberg},\ \bibinfo {year} {1986})\
  \bibinfo {note} {chapter II by J. Leiterer}\BibitemShut {NoStop}%
\bibitem [{\citenamefont {Henkin}\ and\ \citenamefont
  {Leiterer}(1998)}]{Henkin1998}%
  \BibitemOpen
  \bibfield  {author} {\bibinfo {author} {\bibfnamefont {G.}~\bibnamefont
  {Henkin}}\ and\ \bibinfo {author} {\bibfnamefont {J.}~\bibnamefont
  {Leiterer}},\ }\href
  {https://webusers.imj-prg.fr/{~}guennadi.henkin/mathan98.pdf} {\bibfield
  {journal} {\bibinfo  {journal} {Math. Ann}\ }\textbf {\bibinfo {volume}
  {311}},\ \bibinfo {pages} {71} (\bibinfo {year} {1998})}\BibitemShut
  {NoStop}%
\bibitem [{\citenamefont {Marzari}\ and\ \citenamefont
  {Vanderbilt}(1997)}]{Marzari1997}%
  \BibitemOpen
  \bibfield  {author} {\bibinfo {author} {\bibfnamefont {N.}~\bibnamefont
  {Marzari}}\ and\ \bibinfo {author} {\bibfnamefont {D.}~\bibnamefont
  {Vanderbilt}},\ }\href {\doibase 10.1103/PhysRevB.56.12847} {\bibfield
  {journal} {\bibinfo  {journal} {Phys. Rev. B}\ }\textbf {\bibinfo {volume}
  {56}},\ \bibinfo {pages} {12847} (\bibinfo {year} {1997})}\BibitemShut
  {NoStop}%
\bibitem [{\citenamefont {Chen}\ \emph {et~al.}(2013)\citenamefont {Chen},
  \citenamefont {Gu}, \citenamefont {Liu},\ and\ \citenamefont
  {Wen}}]{Chen2013a}%
  \BibitemOpen
  \bibfield  {author} {\bibinfo {author} {\bibfnamefont {X.}~\bibnamefont
  {Chen}}, \bibinfo {author} {\bibfnamefont {Z.-C.}\ \bibnamefont {Gu}},
  \bibinfo {author} {\bibfnamefont {Z.-X.}\ \bibnamefont {Liu}}, \ and\
  \bibinfo {author} {\bibfnamefont {X.-G.}\ \bibnamefont {Wen}},\ }\href
  {\doibase 10.1103/PhysRevB.87.155114} {\bibfield  {journal} {\bibinfo
  {journal} {Phys. Rev. B}\ }\textbf {\bibinfo {volume} {87}},\ \bibinfo
  {pages} {155114} (\bibinfo {year} {2013})}\BibitemShut {NoStop}%
\bibitem [{\citenamefont {Wang}\ \emph {et~al.}(2017)\citenamefont {Wang},
  \citenamefont {Wen},\ and\ \citenamefont {Witten}}]{Wang2017}%
  \BibitemOpen
  \bibfield  {author} {\bibinfo {author} {\bibfnamefont {J.}~\bibnamefont
  {Wang}}, \bibinfo {author} {\bibfnamefont {X.-G.}\ \bibnamefont {Wen}}, \
  and\ \bibinfo {author} {\bibfnamefont {E.}~\bibnamefont {Witten}},\ }\href
  {http://arxiv.org/abs/1705.06728} {\bibfield  {journal} {\bibinfo  {journal}
  {arXiv}\ } (\bibinfo {year} {2017})},\ \Eprint
  {http://arxiv.org/abs/1705.06728} {arXiv:1705.06728} \BibitemShut {NoStop}%
\bibitem [{\citenamefont {Rabson}\ and\ \citenamefont
  {Fisher}(2001)}]{Rabson2001}%
  \BibitemOpen
  \bibfield  {author} {\bibinfo {author} {\bibfnamefont {D.~A.}\ \bibnamefont
  {Rabson}}\ and\ \bibinfo {author} {\bibfnamefont {B.}~\bibnamefont
  {Fisher}},\ }\href {\doibase 10.1103/PhysRevB.65.024201} {\bibfield
  {journal} {\bibinfo  {journal} {Phys. Rev. B}\ }\textbf {\bibinfo {volume}
  {65}},\ \bibinfo {pages} {024201} (\bibinfo {year} {2001})}\BibitemShut
  {NoStop}%
\bibitem [{\citenamefont {Fang}\ and\ \citenamefont
  {Fu}(2015{\natexlab{b}})}]{Fang2015S}%
  \BibitemOpen
  \bibfield  {author} {\bibinfo {author} {\bibfnamefont {C.}~\bibnamefont
  {Fang}}\ and\ \bibinfo {author} {\bibfnamefont {L.}~\bibnamefont {Fu}},\
  }\href {\doibase 10.1103/PhysRevB.91.161105} {\bibfield  {journal} {\bibinfo
  {journal} {Phys. Rev. B}\ }\textbf {\bibinfo {volume} {91}},\ \bibinfo
  {pages} {161105} (\bibinfo {year} {2015}{\natexlab{b}})},\ \bibinfo {note}
  {supplementary material}\BibitemShut {NoStop}%
\bibitem [{\citenamefont {Schnyder}\ \emph {et~al.}(2008)\citenamefont
  {Schnyder}, \citenamefont {Ryu}, \citenamefont {Furusaki},\ and\
  \citenamefont {Ludwig}}]{schnyder_classify3DTIandTSC}%
  \BibitemOpen
  \bibfield  {author} {\bibinfo {author} {\bibfnamefont {A.~P.}\ \bibnamefont
  {Schnyder}}, \bibinfo {author} {\bibfnamefont {S.}~\bibnamefont {Ryu}},
  \bibinfo {author} {\bibfnamefont {A.}~\bibnamefont {Furusaki}}, \ and\
  \bibinfo {author} {\bibfnamefont {A.~W.~W.}\ \bibnamefont {Ludwig}},\
  }\href@noop {} {\bibfield  {journal} {\bibinfo  {journal} {Phys. Rev. B}\
  }\textbf {\bibinfo {volume} {78}},\ \bibinfo {pages} {195125} (\bibinfo
  {year} {2008})}\BibitemShut {NoStop}%
\bibitem [{\citenamefont {Cano}\ \emph {et~al.}(2017)\citenamefont {Cano},
  \citenamefont {Bradlyn}, \citenamefont {Wang}, \citenamefont {Elcoro},
  \citenamefont {Vergniory}, \citenamefont {Felser}, \citenamefont {Aroyo},\
  and\ \citenamefont {Bernevig}}]{Cano2017a}%
  \BibitemOpen
  \bibfield  {author} {\bibinfo {author} {\bibfnamefont {J.}~\bibnamefont
  {Cano}}, \bibinfo {author} {\bibfnamefont {B.}~\bibnamefont {Bradlyn}},
  \bibinfo {author} {\bibfnamefont {Z.}~\bibnamefont {Wang}}, \bibinfo {author}
  {\bibfnamefont {L.}~\bibnamefont {Elcoro}}, \bibinfo {author} {\bibfnamefont
  {M.~G.}\ \bibnamefont {Vergniory}}, \bibinfo {author} {\bibfnamefont
  {C.}~\bibnamefont {Felser}}, \bibinfo {author} {\bibfnamefont {M.~I.}\
  \bibnamefont {Aroyo}}, \ and\ \bibinfo {author} {\bibfnamefont {B.~A.}\
  \bibnamefont {Bernevig}},\ }\href {http://arxiv.org/abs/1711.11045} {\
  (\bibinfo {year} {2017})},\ \Eprint {http://arxiv.org/abs/1711.11045}
  {arXiv:1711.11045} \BibitemShut {NoStop}%
\bibitem [{\citenamefont {Dubail}\ and\ \citenamefont
  {Read}(2015)}]{Dubail2015}%
  \BibitemOpen
  \bibfield  {author} {\bibinfo {author} {\bibfnamefont {J.}~\bibnamefont
  {Dubail}}\ and\ \bibinfo {author} {\bibfnamefont {N.}~\bibnamefont {Read}},\
  }\href {\doibase 10.1103/PhysRevB.92.205307} {\bibfield  {journal} {\bibinfo
  {journal} {Phys. Rev. B}\ }\textbf {\bibinfo {volume} {92}},\ \bibinfo
  {pages} {1} (\bibinfo {year} {2015})}\BibitemShut {NoStop}%
\bibitem [{\citenamefont {H\"ormander}(1989)}]{Hormander1989}%
  \BibitemOpen
  \bibfield  {author} {\bibinfo {author} {\bibfnamefont {L.}~\bibnamefont
  {H\"ormander}},\ }\href@noop {} {\emph {\bibinfo {title} {{An introduction to
  complex analysis in several variables}}}}\ (\bibinfo  {publisher}
  {North-Holland},\ \bibinfo {year} {1989})\BibitemShut {NoStop}%
\bibitem [{\citenamefont {Oka}(1939)}]{Oka1939}%
  \BibitemOpen
  \bibfield  {author} {\bibinfo {author} {\bibfnamefont {K.}~\bibnamefont
  {Oka}},\ }\href@noop {} {\bibfield  {journal} {\bibinfo  {journal} {J. Sc.
  Hiroshima Univ.}\ }\textbf {\bibinfo {volume} {9}},\ \bibinfo {pages} {7}
  (\bibinfo {year} {1939})}\BibitemShut {NoStop}%
\bibitem [{\citenamefont {Huckleberry}(2013)}]{Huckleberry2013}%
  \BibitemOpen
  \bibfield  {author} {\bibinfo {author} {\bibfnamefont {A.}~\bibnamefont
  {Huckleberry}},\ }\href {http://arxiv.org/abs/1303.6933} {\bibfield
  {journal} {\bibinfo  {journal} {arXiv}\ } (\bibinfo {year} {2013})},\ \Eprint
  {http://arxiv.org/abs/1303.6933} {arXiv:1303.6933} \BibitemShut {NoStop}%
\end{thebibliography}%

\end{document}